\documentclass[usenatbib]{mn2e}

\usepackage{graphicx}
\usepackage{epsfig}
\usepackage{subfigure}

\newcommand{\ha}{H$\alpha$} 
\newcommand{\hi}{H{\sc i}} 
\newcommand{\hii}{H{\sc ii}}

\def\vhel{\ifmmode{V_{{\rm HEL}}}\else{$V_{{\rm HEL}}$}\fi}
\def\vsys{\ifmmode{V_{\rm sys}}\else{$V_{\rm sys}$}\fi}
\def\kms{\ifmmode{~{\rm km\,s}^{-1}}\else{~km~s$^{-1}$}\fi}
\def\vlsr{\ifmmode{v_{\rm lsr}}\else{$v_{\rm lsr}$}\fi}

\def\ltsim{\ifmmode\stackrel{<}{_{\sim}}\else$\stackrel{<}{_{\sim}}$\fi}
\def\gtsim{\ifmmode\stackrel{>}{_{\sim}}\else$\stackrel{>}{_{\sim}}$\fi}

\def\reff@jnl#1{{\rm#1\/}}

\def\aj{\reff@jnl{AJ}}                  
\def\araa{\reff@jnl{ARA\&A}}            
\def\apj{\reff@jnl{ApJ}}                
\def\apjl{\reff@jnl{ApJ}}               
\def\apjs{\reff@jnl{ApJS}}              
\def\ao{\reff@jnl{Appl.Optics}}         
\def\apss{\reff@jnl{Ap\&SS}}            
\def\aap{\reff@jnl{A\&A}}               
\def\aapr{\reff@jnl{A\&A~Rev.}}         
\def\aaps{\reff@jnl{A\&AS}}             
\def\azh{\reff@jnl{AZh}}                        
\def\baas{\reff@jnl{BAAS}}              
\def\jrasc{\reff@jnl{JRASC}}            
\def\memras{\reff@jnl{MmRAS}}           
\def\mnras{\reff@jnl{MNRAS}}            
\def\pra{\reff@jnl{Phys.Rev.A}}         
\def\prb{\reff@jnl{Phys.Rev.B}}         
\def\prc{\reff@jnl{Phys.Rev.C}}         
\def\prd{\reff@jnl{Phys.Rev.D}}         
\def\prl{\reff@jnl{Phys.Rev.Lett}}      
\def\pasp{\reff@jnl{PASP}}              
\def\pasj{\reff@jnl{PASJ}}              
\def\qjras{\reff@jnl{QJRAS}}            
\def\skytel{\reff@jnl{S\&T}}            
\def\solphys{\reff@jnl{Solar~Phys.}}    
\def\sovast{\reff@jnl{Soviet~Ast.}}     
 \def\ssr{\reff@jnl{Space~Sci.Rev.}}     
\def\zap{\reff@jnl{ZAp}}                        
\def\nat{\reff@jnl{Nature}}             

\def\LaTeX{L\kern-.36em\raise.3ex\hbox{a}\kern-.15em
    T\kern-.1667em\lower.7ex\hbox{E}\kern-.125emX}

\begin{document}

\title[Diffuse RRL emission on the Galactic plane]{Diffuse RRL emission on the Galactic plane between $\ell=36$\degr and $44$\degr}
\author[M.I.R. Alves et al.]{Marta I. R. Alves,$\!^{1}$\thanks{E-mail:malves@jb.man.ac.uk} Rodney D. Davies,$\!^{1}$ Clive Dickinson,$\!^{1}$ Richard J. Davis,$\!^{1}$\newauthor Robert R. Auld,$\!^{2}$ Mark Calabretta,$\!^{3}$ Lister Staveley-Smith$^{4}$ \\
$^1$Jodrell Bank Centre for Astrophysics, Alan Turing Building, School of Physics and Astronomy, \\
The University of Manchester, Oxford Road, Manchester, M13 9PL, UK \\
$^2$School of Physics \& Astronomy, Cardiff University, Queens Buildings, The Parade, Cardiff, CF24 3AA, UK \\ 
$^3$Australia Telescope National Facility, PO Box 76, Epping, NSW 1710, Australia\\
$^4$International Centre for Radio Astronomy Research, School of Physics, The University of Western Australia,\\
35 Stirling Hwy, Crawley, WA 6009, Australia\\
}

\date{Received **insert**; Accepted **insert**}
       
\pagerange{\pageref{firstpage}--\pageref{lastpage}} 
\pubyear{}

\maketitle
\label{firstpage}


\begin{abstract}

Radio recombination lines (RRLs) can be used to determine the emission measure unambiguously along the Galactic plane.  We use the deep (2100s per beam) HI Parkes Zone of Avoidance survey which includes 3 RRLs (H$166\alpha$, H$167\alpha$ and H$168\alpha$) within its bandwidth.  The region $\ell = 36\degr$ to $44\degr$, $b = -4\degr$ to $+4\degr$ is chosen to include emission from the Local, Sagittarius and Scutum arms.  An $8\degr \times 8\degr$ data cube centred at $(\ell, b) = (40\degr , 0\degr)$ is constructed of RRL spectra with velocity and spatial resolution of 27$\kms$ and 15.5 arcmin, respectively.  Well-known \hii~regions are identified as well as the diffuse RRL emission on the Galactic plane.  A Galactic latitude section of the integrated RRL emission across the Galactic plane delineates the brightness temperature ($T_{b}$) distribution which has a half-power width in latitude of $\simeq 1\fdg5$. 

A value of the electron temperature $T_{e} \simeq 8000$ K is derived from a comparison with the WMAP free-free MEM model. The $T_{b}$ distribution from the present RRL data is combined with the WMAP 5-yr data to derive the anomalous dust on the Galactic ridge.

In this paper we demonstrate that diffuse ionized emission on the Galactic ridge can be recovered using RRLs from the ZOA survey. This method is therefore able to complement the \ha~data at low Galactic latitudes, to enable an all-sky free-free template to be derived.

\end{abstract}



\setcounter{figure}{0}

\section{INTRODUCTION}
\label{sec:introduction}

Galactic free-free emission is a principal foreground contaminant of the Cosmic
Microwave Background (CMB).  It becomes a major component near the
Galactic plane where it is produced in the gas layer ionized by radiation from 
recently formed stars.  A true all-sky free-free foreground template
needs to include this low Galactic latitude component if the CMB power
spectrum is to be correctly evaluated at low values of angular
frequency, $\ell$, where there is currently significant interest.

The diffuse electron gas along the Galactic plane is also of intrinsic
interest.  It is an indicator of current star formation throughout the
Galaxy and complements the data available from catalogues of
\hii~regions \citep{2003A&A...397..213P} which represent localized areas of
star formation. There is considerable interest in the relationship between this narrow WIM
distribution with a scale height of $z \sim $ 150 pc and the thick disc component with
$z \sim $ 1000 pc responsible for pulsar dispersion measures \citep{2008PASA...25..184G}.
Issues of concern are the clumpiness (filling factor), the scale height, and
the electron temperature as well as the transport of ionizing radiation from
the hot stars in the narrow distribution to the broad distribution.

One of the direct ways of obtaining a free-free emission template is
to use large-area \ha~surveys.  These include the WHAM survey \citep{2003ApJS..149..405H} 
which uses Fabry-Perot spectroscopy on an angular scale
of $\approx 1\degr$ covering the northern sky (Dec~$> -30\degr$). 
The WHAM instrument is now installed at the Cerro Tololo Interamerican Observatory (CTIO) for the southern counterpart of the survey. The SHASSA survey \citep{2001PASP..113.1326G} employs
a filter system with an angular resolution of $\approx 1$~arcmin and covers
the southern sky;  it is however limited in sensitivity due to
geocoronal emission and is insensitive to large angular scales
($>10\degr$) due to the field-of-view. The VTSS survey \citep*{1998PASA...15..147D} 
uses a similar filter approach on the northern sky on a 2
arcmin scale.  Before the data from these surveys can be used to
derive a free-free template, the absorption by interstellar dust must
be taken into account;  this correction is significant at intermediate
and low latitudes. \citet*{2003MNRAS.341..369D} have used the
\citet*{1998ApJ...500..525S} dust map derived from the IRAS and
DIRBE $100~\mu$m all-sky surveys. At $|b| \ltsim 5\degr$ this correction
breaks down and accordingly the \ha~surveys cannot be used for this
purpose along the Galactic plane.  The construction of a free-free
template at latitudes where absorption becomes significant can be
attempted at GHz radio frequencies by exploiting the difference of
temperature spectral index between free-free ($\beta = -2.1$) and
synchrotron ($\beta \approx -2.7$) where $T \propto \nu^{\beta}$.  If
3 frequencies are available the free-free temperature can be derived
along with the synchrotron temperature and spectral index;  this
situation is somewhat complicated by the variation of spectral index
with position and frequency. 

Radio recombination lines (RRLs) in contrast to \ha~and free-free
emission can be used to obtain the emission measure (${\rm EM} = \int{n_{e}^2 dl}$)
directly with no absorption or contamination.  The lines themselves
and the ionized medium in which they are immersed are optically thin
at the frequencies used for this work.  However, in order to convert
the RRL line integral to a free-free emission brightness temperature a knowledge
of the electron temperature $T_{e}$ is required;  there are a number of
routes to obtaining $T_{e}$.

This paper describes an exploration of the RRL approach to obtaining a
free-free template of a section of the Galactic plane at
$\ell=36\degr - 44\degr$, $|b| < 4\degr$ in data from the HI Zone of 
Avoidance Survey using the Parkes 13-beam H-line receiver
system. Section~\ref{sec:rrl} summarizes RRL theory and examines
previous RRL surveys which were of high sensitivity but limited area.
In Section~\ref{sec:rrl_survey} we give the survey parameters and the
criteria for selecting the longitude range for study.  The data reduction 
and the RRL map derived in this study are described in
Section~\ref{sec:data_analysis}. Sections \ref{sec:free-free_from_rrl}
and \ref{sec:latitude_distribution} convert the RRL line integrals
into $T_{b}$ and compare them with the 21 cm continuum derived from this 
survey, as well as with \ha~and WMAP data.


\section{Radio Recombination Lines}
\label{sec:rrl}

The RRL emission from a diffuse ionized hydrogen gas can be expressed
in terms of the integral over spectral line temperature as
\begin{equation}
\int T_{L} d \nu = 1.92 \times 10^{3}  T_{e}^{-1.5}  {\rm EM} \label{eq:1}
\end{equation}

where $\nu$ is the frequency (kHz), $T_{e}$ is the electron temperature (K) and the
emission measure EM is in cm$^{-6}$pc.  The corresponding continuum
emission brightness temperature is
\begin{equation}
T_{C} = 8.235 \times 10^{-2} a(T_{e})  T_{e}^{-0.35} \nu^{-2.1}_{\rm GHz}
 (1+0.08)  {21.9969\rm EM} \label{eq:2}
\end{equation}

where $a(T_{e})$ is a slowly varying function of temperature and $\nu$
is the frequency in GHz \citep{1967ApJ...147..471M}.  The (1+0.08) term represents the additional
contribution to $T_{C}$ from helium.  These equations lead to an
expression for the ratio of the line integral to the continuum
\begin{equation}
\frac{\int T_{L} dV}{T_{C}} = 6.985 \times 10^{3} \frac{1}{a(T_{e})}
 \frac{1}{n(He)/(1 + n(H))}  T_{e}^{-1.15}  \nu^{1.1}_{\rm GHz} \label{eq:3}
\end{equation}

where $V$ is in \kms~\citep{2004tra..book.....R}.

Most RRL surveys have concentrated on individual \hii~regions for
which an adequate continuum antenna temperature is achievable.  The
\hii~region catalogue of \citet{2003A&A...397..213P} lists $\simeq 800$ sources
with RRL data \citep*{2004MNRAS.347..237P}.

At frequencies near 1~GHz the recombination lines are relatively weak;
for a typical line width of 25~\kms~and $T_{e} \approx 10^{4}$~K, 
$T_{L}/T_{C} \approx 1$ per cent and increases as $\nu^{1.1}$.  The
thermal ridge of the Galactic plane has a brightness temperature of
$5-10$~K at 1~GHz (see Section~\ref{sec:latitude_distribution}) in the
area we have chosen to study and accordingly the peak diffuse RRL
emission temperature will be $50-100$~mK and correspondingly less if
the emission is spread over a larger velocity range due to Galactic
differential rotation for example.  If we are to determine the
latitude width of the free-free emission, a sensitivity approaching a few mK is required.  This implies integration times of $\sim 1$
hour with modern receiver systems at $\nu \sim 1$ GHz.

Previous surveys of the diffuse Galactic free-free emission have of
necessity been restricted to observations of grids of points along the
Galactic plane, mainly on the Galactic ridge ($b=0\degr$).
Table~\ref{tab:rrl_surveys} lists the substantive previous surveys,
all in the H$166\alpha$ (1424.734~MHz) or H$157\alpha$ (1683.2~MHz)
lines. The beamwidth used and survey coordinates are given.  An
examination of these surveys shows line brightness temperatures of
~50-100 mK in areas away from strong \hii~regions.

\begin{table*}
\centering
 \caption{RRL surveys of diffuse emission on the Galactic plane. \label{tab:rrl_surveys}}
  \begin{tabular}{lcccc}
    \hline
Authors &Transition &Frequency &Beamwidth &Longitudes~$\ell$   \\
        &           &(MHz)     &(arcmin)  &                 \\   \hline

\citet{1972ApJ...176..587G}   &H$157\alpha$    &1683.2    &33      &$9\degr \rightarrow 130\degr$  \\
\citet{1976MNRAS.176..547H}   &H$166\alpha$    &1424.734   &$31 \times 33$      &$5\degr \rightarrow 70\degr$  \\
\citet{1976ApJ...209..429L}   &H$166\alpha$    &1424.734   &21      &$358\degr \rightarrow  50\fdg5$  \\
\hline
 \end{tabular}
\end{table*}

Multibeam, low-noise systems are required to make a significant improvement to the situation outlined in Table~\ref{tab:rrl_surveys}.


\section{The RRL survey}
\label{sec:rrl_survey}

A sensitive multibeam survey for RRL emission on a section of the
Galactic plane is a by-product of the Parkes \hi~All-Sky Survey
(HIPASS, \citealt{1996PASA...13..243S}) and associated Jodrell Bank Observatory \hi~All-Sky Survey
(HIJASS).  The aim was to detect the HI-emitting galaxies in the local
Universe using a bandwidth of 64~MHz corresponding to redshifted velocities of $-1,280$ and $12,700$~\kms. In this
velocity range is Galactic \hi~and HVC emission at
near-Galactic velocities \citep{2002AJ....123..873P}.  In addition there are
3 Galactic RRL lines (H$166\alpha$, H$167\alpha$, and H$168\alpha$) in
this frequency range.

HIPASS observations are taken by scanning the telescope in declination strips of 8\degr~length, separated by 7 arcmin in right ascension. The footprint of the 13 beam receiver on the sky is $1\fdg7$ and it scans at 1\degr/min, recording the signal every 5 seconds. The integration time is 450s/beam \citep{2001MNRAS.322..486B}.

Of particular interest for the study of the necessarily weak RRL emission from the Galactic plane is the Zone of Avoidance (ZOA) \hi~Survey associated with HIPASS. This searched for galaxies behind the Galactic plane and had a factor of 5 longer integration time than the HIPASS and HIJASS surveys \citep{2005AJ....129..220D}.  The observing parameters used for the HIPASS ZOA survey which are relevant to the RRL study are given in Table~\ref{tab:zoa}. The ZOA survey was scanned in Galactic coordinates, along strips of constant latitude, separated by 1.4 arcmin.
The integration time of 2100 seconds per beam leads to an r.m.s. sensitivity per beam per channel of 4.6 mK in brightness temperature. 
We adopt the brightness temperature conversion factor of 0.8 K/Jy, assuming a beam size of 15.5 arcmin, as given by \citet{2002AJ....123..873P}.
When the 3 RRLs are co-added the r.m.s. sensitivity is 2.6 mK which is the level required to map the Galactic ridge RRL emission.  The challenge is to achieve this sensitivity in the presence of baseline effects and interference as we shall see below.

\begin{table}
\centering
 \caption{Observing parameters of the HIPASS ZOA survey as applied to Galactic RRLs.\label{tab:zoa}}
  \begin{tabular}{lc}
    \hline
Frequency coverage			&1362 to 1426 MHz             \\
RRL lines included:                      &                             \\
~~~~~H$166\alpha$ 			&1424.735 MHz              \\
~~~~~H$167\alpha$ 			&1399.368 MHz             \\
~~~~~H$168\alpha$ 			&1374.601 MHz             \\
Longitude coverage		        &$\ell = 196\degr-52\degr$ \\
Latitude coverage			&$b = -4\degr$ to $+4\degr$     \\
Beamwidth (FWHP)			&14.3 arcmin \\
Beamwidth (after gridding)		&15.5 arcmin \\
Grid pixel size				&4 arcmin    \\
Channel separation$^{\rm a}$		&13.2~\kms   \\
Velocity resolution (Tukey)		&18.0~\kms    \\
Velocity resolution (Tukey+Hanning)	&27.0~\kms    \\
Integration time per beam		&2100~s       \\
r.m.s. noise$^{\rm b,c}$ per channel		&5.7~mJy/beam \\
\hline
\multicolumn{2}{|l|}{(a) corresponds to 64 kHz at 1420.6 MHz; 1000 channels in} \\
\multicolumn{2}{|l|}{64 MHz (b) the equivalent $T_{b}$ in a beam is 4.6 mK} \\
\multicolumn{2}{|l|}{(0.8 K/Jy) (c) each beam includes the sum of 2 orthogonal} \\
\multicolumn{2}{|l|}{linear polarizations.} \\
\end{tabular}
\end{table}

\begin{figure}
\centering
\vspace{-0.3cm}
\includegraphics[scale=0.23]{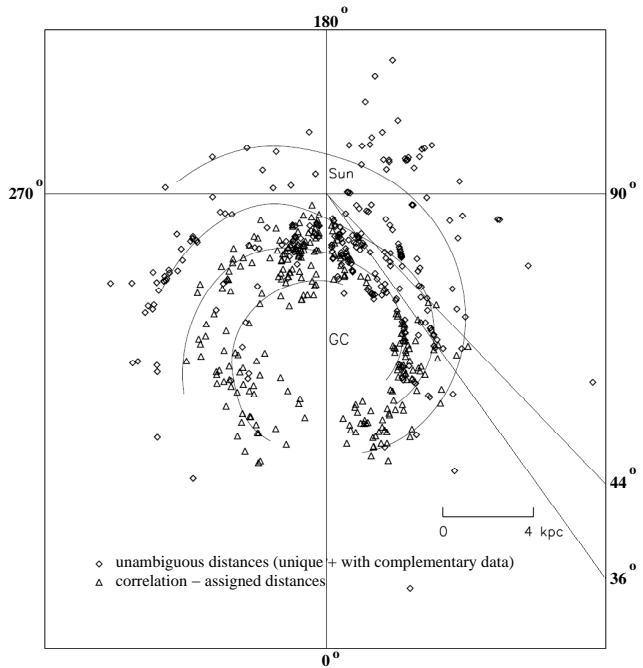}
\caption{Galactic spiral arms outlined in HII regions \citep{2004MNRAS.347..237P} with the $\ell=36\degr$ to $44\degr$ sector observed in the RRL data cube. 550 \hii~regions from the \citet{2003A&A...397..213P} catalogue are shown based on the spiral arm model by \citet{1993ApJ...411..674T}.}
\label{fig1}
\end{figure}

The angular resolution of the survey ($\sim 15$~arcmin) is adequate to
investigate the likely structure of the diffuse free-free emission.
The z-distribution of \hii~regions inside the solar circle has $\sigma \sim
40-50$~pc \citep{2004MNRAS.347..237P}, somewhat greater than that of OB
stars \citep{2000A&A...358..521B} but similar to that of the CO layer;
\citet{1991ApJ...372L..17R} has estimated $\sigma = 70$~pc for a thin disc of
\hii~from pulsar dispersion measures. A value of $\sigma = 50$~pc
(a FWHP of $\sim 120$~pc) corresponds to $\simeq 1\degr$ at 3~kpc the
distance of the Sagittarius arm or half these values in the Scutum arm
at the longitudes to be studied here.

The velocity resolution of the survey, even with additional Hanning smoothing, (see Section \ref{sec:data_analysis})
is adequate to resolve the emission from the
local Orion arm ($V = 0~\kms$), the Sagittarius ($40~\kms$) and the
Scutum ($80~\kms$) arms.  However for the purposes of mapping the
diffuse electron gas we are interested only in the line integral over
the emission velocity range 0 to 100~\kms~and we use the integral over
$-20$ to $+120\kms$~in order to take account of any velocity spread
within the arms.

The ZOA survey comprises $8\degr \times 8\degr \times 64$~MHz
data cubes;  each cube covers $8\degr$ in Galactic longitude and $\pm
4\degr$ in Galactic latitude. For this initial analysis we have chosen
the data cube covering the longitude range $36\degr$ to $44\degr$
which includes emission from the Local, Sagittarius and Scutum arms.
The field centre is $(\ell,b) = (40\degr, 0\degr)$. Fig. \ref{fig1} is a map of the \hii~regions distribution in the Galaxy, it shows the geometry of the intersection of the survey area with the 3 arms.
Clearly the major contribution to the RRL emission will be from the
Sagittarius arm which is sampled at distances of 3 to 7~kpc.  The line
of sight is tangent to the Scutum arm at $\sim 6$~kpc.

\section{Analysis of the data cube centred on $(\ell$, \lowercase{b}) = (40\degr, 0\degr)}
\label{sec:data_analysis}

\begin{figure*}
\centering
\includegraphics[scale=0.5]{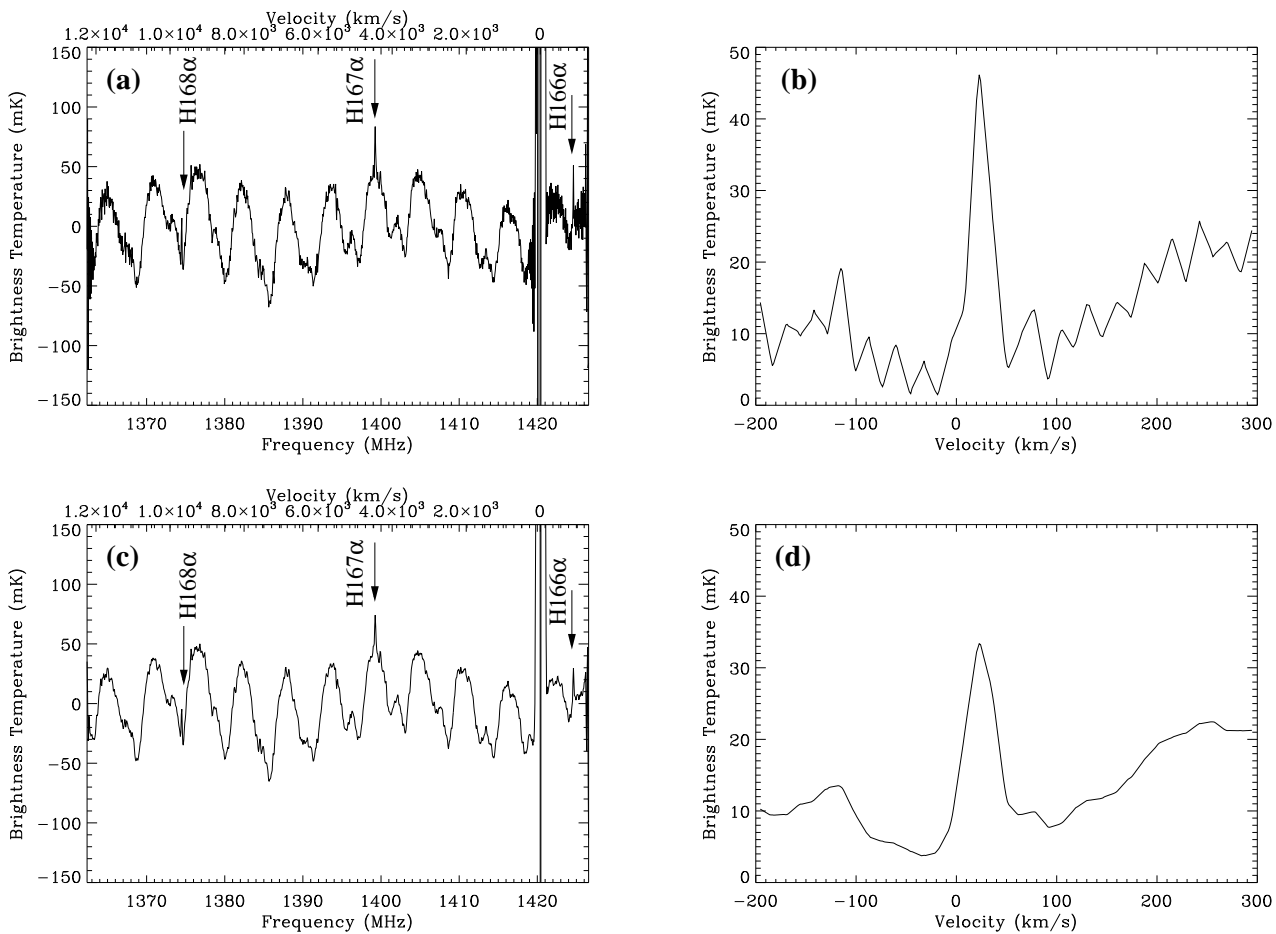}
\caption{Spectrum (left) and stacked line (right) for the \hii~region W45, taking a $12 \times 12$ arcmin$^{2}$ average spectrum. Figs. (a) and (b) are for the Tukey smoothed spectrum and Figs. (c) and (d) for the Tukey+Hanning. The three RRLs are added to produce the final line, with no baseline correction. The 5.7 MHz period sinusoidal ripple caused by the continuum sources is visible in the spectra of Figs. (a) and (c) (see text).}
\label{fig2}
\end{figure*}

\subsection{Data Reduction}
\label{sec:data_red}

Data reduction largely follows the procedures
given in \citet{2001MNRAS.322..486B} using the package {\sc livedata} within the {\sc aips++} environment, which
corrects spectra for bandpass effects and calibrates them.  Standard HIPASS bandpass
correction is performed by dividing the target spectrum by a reference
off-source spectrum which represents the underlying spectral shape as
well as the spectral shape of the receivers, ground pickup and sky
radiation.  The reference spectrum for each receiver is obtained
before and after each target spectrum as the telescope scans the sky;
the median value at each frequency is used to generate a robust
reference spectrum. 

Since the diffuse Galactic RRL emission is a narrow band extended
along the Galactic plane, the more appropriate {\sc minmed5} analysis method is applied.
This method was used successfully by \citet{2002AJ....123..873P} for HVCs.
We recover extended emission in the ZOA data by dividing the 8\degr~scan into 5 sections, then 
finding the median flux density in each section and subsequently using the minimum of the 5 
values as the bandpass correction.
This procedure largely increases the sensitivity of the 
data to large scale structure without substantial loss of flux density, except when
the emission fills the entire 8\degr~scan, as we shall see in Section \ref{sec:tb_distribution}.

The 3-D $\ell$-$b$-velocity cube of bandpass corrected spectra are
generated from the median spectra within 6 arcmin of each pixel point
using the package {\sc gridzilla}. As a result of applying this top-hat 
median gridding, the beam size increases to 15.5 arcmin \citep{2005AJ....129..220D}. The resulting $8\degr \times 8\degr$ data cube has 4 arcmin square pixels and an assigned LSR velocity relative to HI for each of the 1024 velocity channels; these are converted to LSR velocities for the three RRLs with the frequencies given in Table \ref{tab:zoa}.

\begin{figure*}
\centering
\includegraphics[scale=0.45]{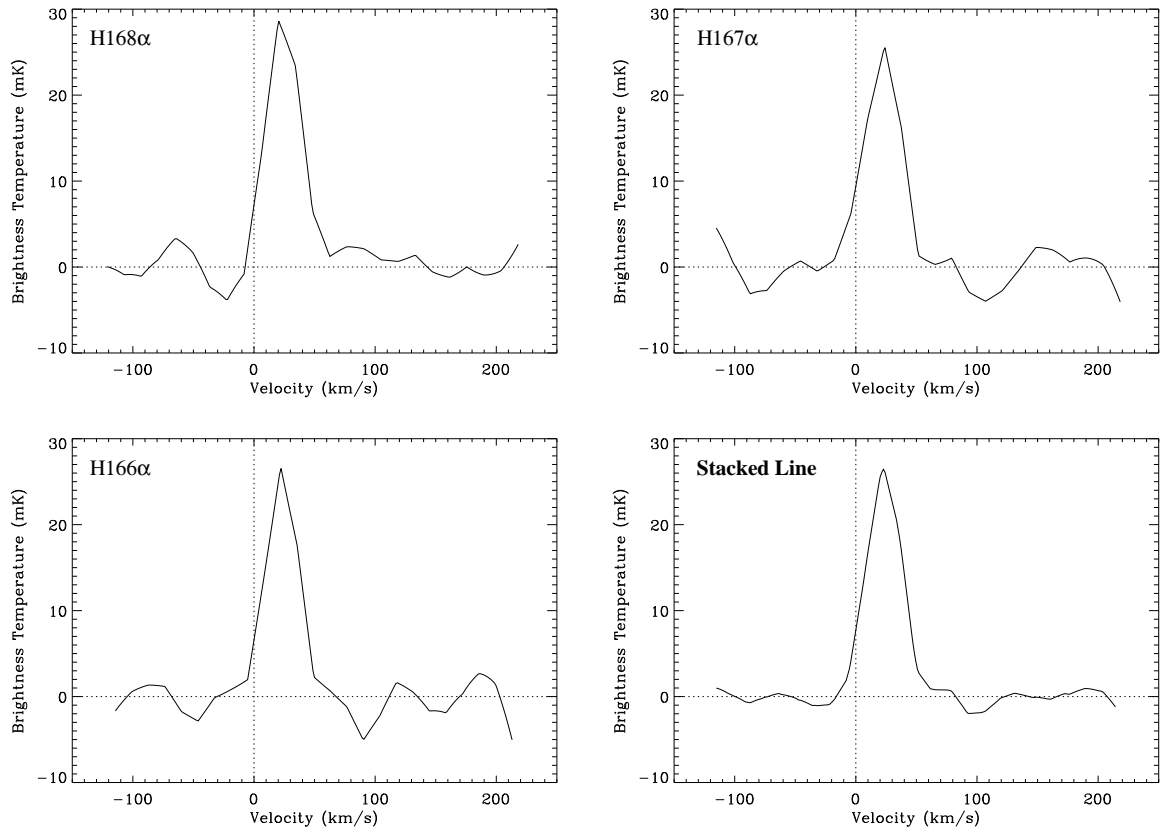}
\caption{The baseline corrected RRLs plus the stacked line for the \hii~region W45, taking a $12 \times 12$ arcmin$^{2}$ average spectrum. The spectrum has been Tukey+Hanning smoothed.}
\label{fig3}
\end{figure*}

\subsection{Standing wave mitigation}
\label{sec:standing_waves}

Radio recombination line studies suffer from the effects of standing waves produced when the continuum waves from the target \hii~region interfere with a fraction of the radiation which reaches the focus along separate paths. Although this extra signal is of the order of a few percent of the main signal, it can be comparable with the intensity of the RRL itself, which in the case of the lines at $\sim$1400 MHz is $\sim$1 \% as mentioned above. Most of the interfering signal results from standing waves between the focus area (focus box and supports) and the apex of the antenna. Similar standing waves are produced when solar radiation scatters into the focus area. Even during the night, scattered radiation can come from Galactic structures adjacent to the \hii~region under investigation. This is the case in the present study of a bright section of the Galactic ridge. Similarly, radiation from the ground can find its way into the focus area. All these sources contribute to the standing waves; they will vary with time of day, the azimuth and elevation and the position in the celestial sky. There is not a unique baseline correction which can be applied to all positions in the present data cube. We now describe our method of dealing with this problem.

Fig. \ref{fig2} shows the spectrum from the data cube at the position of W45 $(\ell,b) = (40\fdg5,+2\fdg5)$, a slightly extended \hii~region with a peak $T_{a} \sim 10$ K. This spectrum is the result of observational data from different focal horns taken at different times. The standing wave pattern is periodic with a frequency of 5.7 MHz and an amplitude of $T_{b}=120$ mK ($T_{a}=80$ mK). This is the frequency expected for standing waves between the focus and the apex of a 64-m antenna with f/D = 0.4. There is additional fine structure in the pattern due to longer interference paths. This fine structure does not repeat exactly over the 64 MHz bandwidth.

In Fig. \ref{fig2}, the 3 RRLs can be clearly seen as narrow lines at their rest velocities with $T_{L} = 30$ mK and are distinguishable from the standing wave pattern. In addition there is higher frequency Gibbs ``ringing'' in the spectrum near the edge of the band and surrounding the strong 1420.406 line of HI. The Tukey smoothing applied in the basic analysis package does not fully remove this digital sampling ringing. Fig. \ref{fig2} shows that the residual Gibbs ringing can be effectively removed by applying an additional process of Hanning smoothing; the price to be paid is a broadening of the spectrum from a half-width of 18 to 27~\kms.

The final step is to correct each RRL for baselines and then to shift and add them, at each position.
The baseline is removed by fitting a polynomial to each line, excluding the velocity range (0 to 100 km/s) where we expect the RRL gas to emit, in this longitude range (Section \ref{sec:rrl_survey}).
Fig. \ref{fig3} illustrates the improvement in the baseline of the spectrum between the original 3 spectra and the added spectrum for W45. The different baseline variations in each of the spectra act effectively like noise and are reduced by nearly a factor of 2 in the addition. It was also necessary to check each spectrum for narrow features either of standing wave or RF interference origin before final acceptance. 

\begin{figure*}
\centering
\includegraphics[scale=0.8]{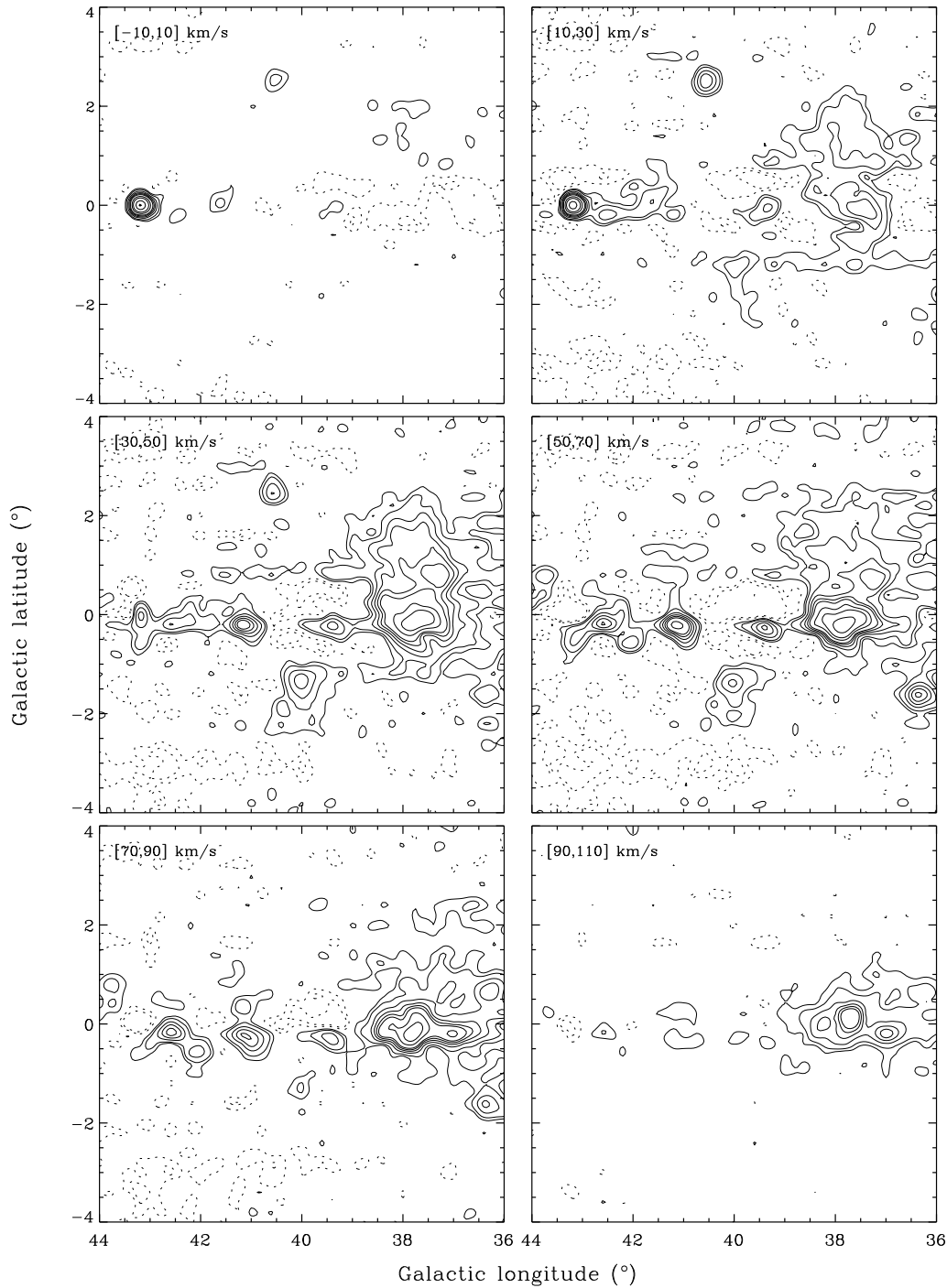}
\caption{Maps of RRL emission centred at 0, 20, 40, 60, 80 and 100~\kms, integrated over 20~\kms, at a resolution of 15.5 arcmin. Contours are given at -2, -1, 1, 2, 4, 6, 8, 10, 15, 20, 30, 40, 50, 60, 70, 80 and 90 per cent of 5.29 K.\kms. The two negative contours are dotted.}
\label{fig4}
\end{figure*}

\begin{figure*}
\centering
\includegraphics[scale=0.5]{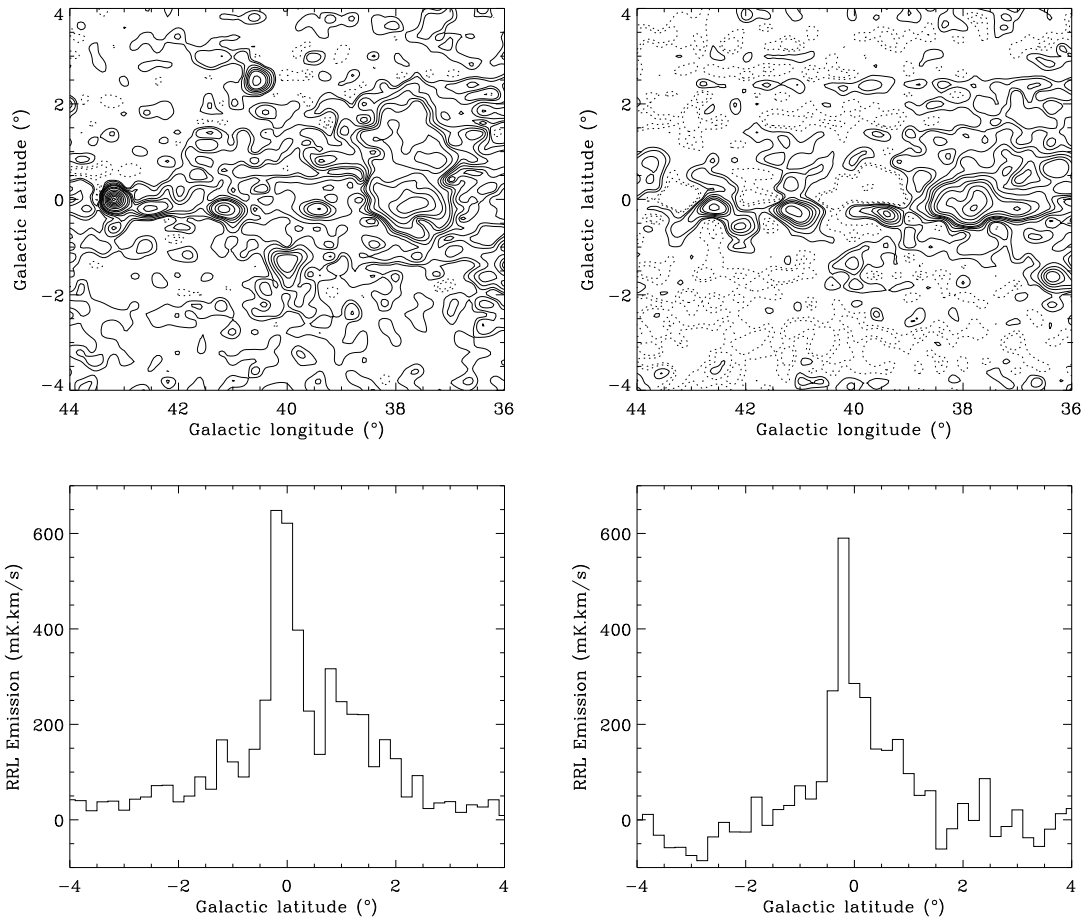}
\caption{Integrated RRL emission for two velocity ranges (top) and the corresponding latitude profiles (bottom). The left and right panels correspond to a line integral from -20 to 60\kms~and from 60 to 120\kms, respectively. The contours are the same as in Fig. \ref{fig4}.}
\label{fig5}
\end{figure*}

\begin{figure*}
\centering
\includegraphics[scale=0.9]{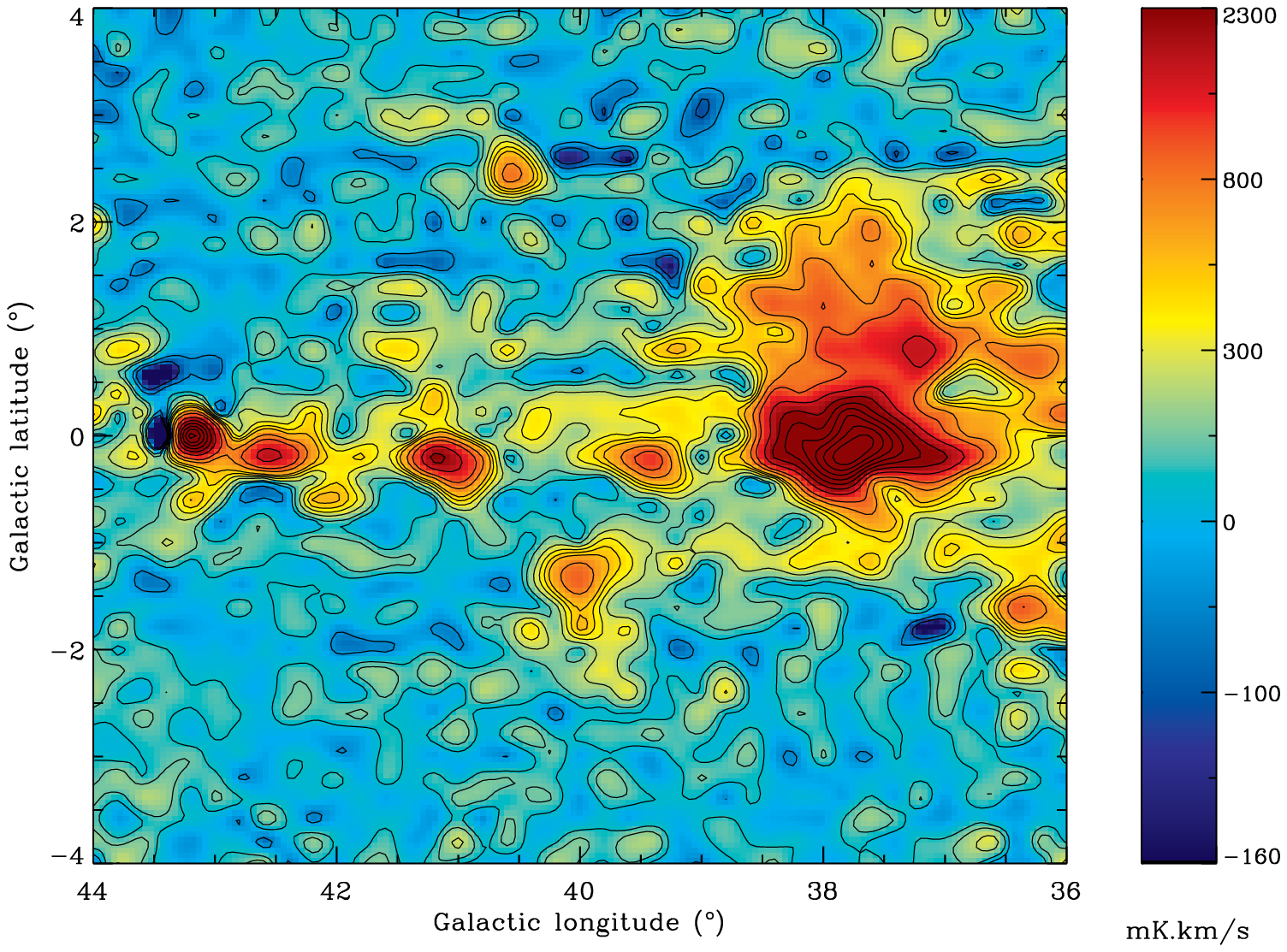}
\caption{Map of the total integrated RRL emission (in mK.kms$^{-1}$) at a resolution of 15.5 arcmin. The integration is made between -20 and 120~\kms. The contours are the same as in Fig. \ref{fig4}, where the value 5.29 K.\kms~is the peak of the line integral in this map.}
\label{fig6}
\end{figure*}

\subsection{The distribution of RRL emission at $\ell=40\degr$}
\label{sec:rrl_dist}

The final RRL data cube shows high significance detection of RRL emission from
individual \hii~regions and from diffuse gas along the Galactic plane.
Fig. \ref{fig4} plots the RRL emission at 6 velocities (0, 20, 40, 60, 80,
100~\kms), integrated over 20~\kms. Individual \hii~regions can be identified on the different maps. Table~\ref{tab:HIIreg} lists known \hii~regions and supernovae in the region $\ell=36\degr$ to $44\degr$.
At the lowest velocities, the emission is broader, since it maps the Local and Sagittarius arms. At the highest velocities ($>$ 60 \kms) the emission is more confined to the Galactic plane and is stronger at low longitudes where the Scutum arm comes in. W49A, $(\ell,b) = (43\fdg2,0\fdg0)$, is the most prominent \hii~region of the field and is one of the strongest \hii~regions of our Galaxy. It is stronger in the first two panels of Fig. \ref{fig4} because its LSR velocity is 7.7\kms, which places it on the far side of our own spiral arm, at $\sim 11.9$ kpc from the Sun.
W45 is also detected at low velocities, V$_{LSR} = 23\kms$. At a latitude of $2\fdg5$ above the plane and a distance of $\sim 1.6$ kpc from the Sun, this \hii~region is in the Local arm.
The extended feature on the plane around $\ell=38\degr$ is a group of \hii~regions with LSR velocities from 50 to 90\kms and is thus in the Scutum arm. The extension to positive latitudes is due to the contribution of some \hii~regions with velocities around 40\kms, and therefore belongs to the Sagittarius spiral arm.

Fig. \ref{fig5} shows two maps of the RRL emission integrated from -20 to 60\kms~and from 60 to 120\kms, and also the corresponding latitude profiles, averaged over $\ell=36\degr$ to $44\degr$. 
Comparing Figs.\ref{fig5} (c) and (d), we can see the broader distribution for low velocity emission with a 
FWHP of $\simeq1\fdg5$, whereas the higher velocity emission has a FWHP of about half that value.

Fig. \ref{fig6} gives the total RRL emission integrated from $-20$ to $120~\kms$. This map shows a clear detection of the diffuse ionized gas around the Galactic plane, as well as the individual \hii~regions. The rms is 70 mK.\kms. The low-level striping in longitude is apparent, due to the telescope scanning strategy and the variation of bandpass correction between adjacent scans. The stripes are at a 2.5 $\sigma$ level, so they only affect the low flux density regions. Some regions near the Galactic plane show sidelobe artifacts due to the emission filling the entire scan (see Section \ref{sec:signalloss}). The biggest negative (4 $\sigma$) is next to one of the strongest \hii~regions in the map, W49.

\section{Estimation of the expected free-free emission from the RRL data}
\label{sec:free-free_from_rrl}

\subsection{The Brightness Temperature ($T_{b}$)}
\label{sec:em}

In order to obtain a value of the $T_{b}$ from the RRL line integral it
is necessary to have a reliable estimate of the electron temperature,
$T_{e}$.  The temperature dependence of $T_{b}$ can be found combining equations (\ref{eq:1}) and (\ref{eq:2}) and is of the form
\begin{equation}
T_{b} = 4.289 \times 10^{-5} a(T_{e}) . T_{e}^{1.15} \nu^{-2.1}_{\rm GHz}
. (1+0.08) . \int T_{L}d\nu
\label{eq:4}
\end{equation}
where $\nu$ is in kHz.

The Emission Measure is derived from equation (\ref{eq:1}). If we adopt the frequency of the central RRL used in this study, namely H$167\alpha$ the effective frequency is 1.40~GHz and we have
\begin{equation}
{\rm EM} = 2.432 \times 10^{-3}  T_{e}^{1.5} . \int T_{L}dV  \label{eq:5}
\end{equation}
where $V$ is in~\kms. 

We must now choose a value of $T_{e}$ that is appropriate for the
diffuse Galactic plane emission in the longitude range of the present
study.  Extensive surveys of RRLs are available which give $T_{e}$
values for individual \hii~regions.  These show a fall in $T_{e}$
towards the Galactic centre due to the greater cooling of the
\hii~region and the stellar temperature of the higher metal content in
the inner Galaxy.  The targeted survey of \citet{1983MNRAS.204...53S} for 67
\hii~regions gave
\begin{equation}
T_{e}~ [{\rm K}] = (3150 \pm 110) + (433 \pm 40)R \label{eq:6}
\end{equation}
where $R$ is the galactocentric distance in kpc;  $R_{0}$ was assumed
to be 8.5~kpc.  \citet{2003A&A...397..213P} used all the published data
for 404 \hii~regions with reliable galactocentric distances and found
\begin{equation}
T_{e}~ [{\rm K}] = (4166 \pm 124) + (314 \pm 20)R \label{eq:7}
\end{equation}
which is in general agreement with the Shaver et al. result.  Any
differences may be due to the larger number of weaker \hii~regions in
the Paladini et al. sample.

We can apply the above relationships to the $\ell = 36\degr - 44\degr$
RRL data cube and derive likely values of $T_{e}$.  Table~\ref{tab:Te}
summarizes the $T_{e}$ values from equations (\ref{eq:6}) and
(\ref{eq:7}) for the near and far galactocentric radii of the Local,
Sagittarius and Scutum arms.  These radii are derived from the spiral
pattern for \hii~regions given in \citet{2004MNRAS.347..237P}.  $T_{e}$
ranges from 6700~K for the Local arm to 5500~K for the Scutum arm.

\begin{table}
\centering
 \caption{Electron temperatures from equations (\ref{eq:6}) and (\ref{eq:7}) for the mean of the near and far galactocentric radii of the line of sight intersection with the Local, Sagittarius and Scutum arms. \label{tab:Te}}
  \begin{tabular}{lccccc}
    \hline
Arm	     &$R_{\rm near}$ &$R_{\rm far}$  &$R_{\rm mean}$  &$T_{e}$~(K)  &$T_{e}$~(K) \\
             &(kpc)    &(kpc)    &(kpc)   &(Shaver)  &(Paladini) \\ \hline
Local        &8.5      &7.5      &8.0     &6610 &6680 \\
Sagittarius  &6.5      &6.0      &6.25    &5860 &6130 \\ 
Scutum       &5.0      &5.0      &5.0     &5310 &5740 \\ \hline
 \end{tabular}
\end{table}

We should discuss whether the electron temperature for individual
\hii~regions to which the equations (6) and (7) refer also applies to
the diffuse \hii~gas.  Since the latitude distribution of the
\hii~regions and the diffuse \hii~are similar as shown in
Section~\ref{sec:latitude_distribution}, the metal abundance
responsible for cooling is expected to be the same.  The main difference will be
the effective temperature of the radiation field which will be lower,
arguing for a lower $T_{e}$. Direct measurements of emission lines of
intermediate latitude ionized gas give values of $T_{e} \sim 8000$~K
\citep{1985ApJ...294..256R} significantly higher than suggested by
Table~\ref{tab:Te}.  
We will adopt a value of $T_{e} = 7000$~K for the present discussion
and note that 1.0 K.\kms~is equivalent to 2.8 K of brightness temperature at 1.4 GHz, 
with $T_{b}$ proportional to $T_{e}^{1.15}$, while $EM \propto T{e}^{1.5}$.

\subsection{The distribution of $T_{b}$ in the $\ell = 40\degr$ data cube}
\label{sec:tb_distribution}

\begin{figure*}
\hspace{1.0cm}
\centering
\includegraphics[scale=0.8]{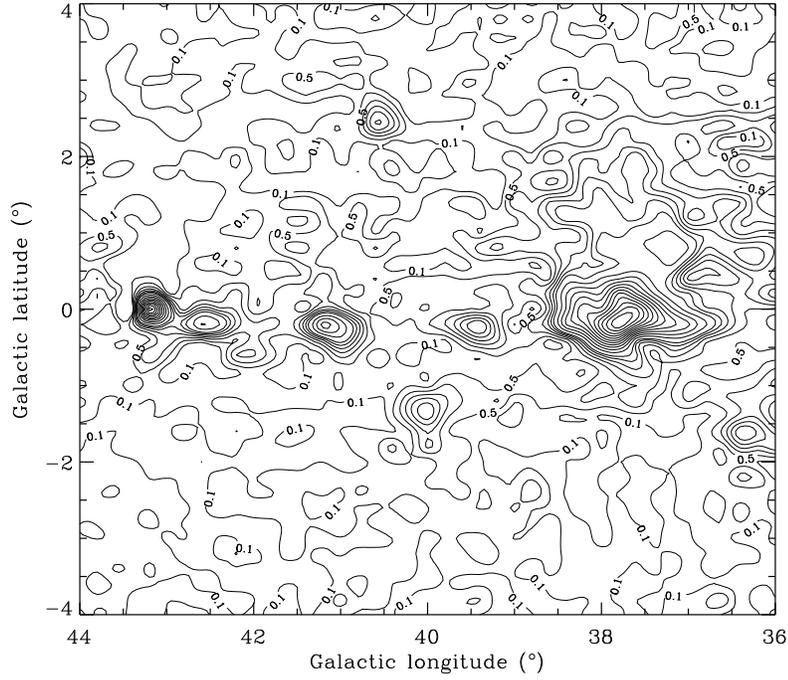}
\caption{Map of brightness temperature at 1.4 GHz estimated from the RRL line integral over the data cube. Contours are given at 0.1 and 0.5 (labeled), 1, 1.5, 2, 2.5, 3 K and then every 1 K until 19 K. The resolution of the map is 15.5 arcmin. Note that 2.8 K ($T_{b}$) corresponds to 1.0 K.\kms, for $T_{e} = 7000$ K.}
\label{fig7}
\end{figure*}

\begin{figure*}
\centering
\hspace{1.0cm}
\includegraphics[scale=0.8]{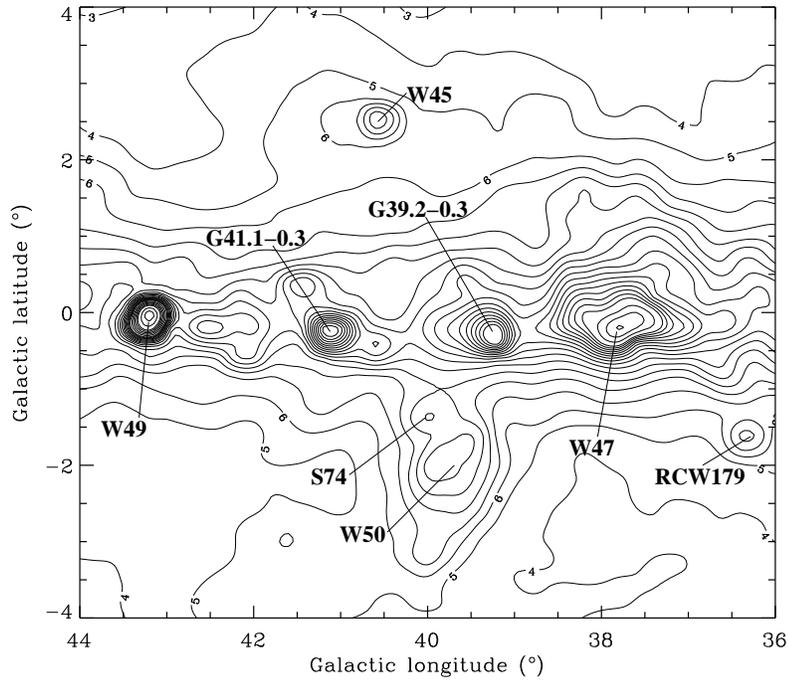}
\caption{Map of 1.4 GHz continuum emission for the region $\ell = 36\degr - 44\degr$, $|b| < 4\degr$, from the combination of ZOA and HIPASS surveys. The contours are given at every 1 K from 3 to 20 K, every 2 K from 20 to 30 K and at 35, 40, 45, 50 and 55 K. The first four contours are labelled on the map. Some of the sources in Table \ref{tab:HIIreg} are labelled. The resolution is 15.5 arcmin.}
\label{fig8}
\end{figure*}

\subsubsection{Comparison with the 1.4 GHz continuum}
\label{sec:tb_and_cont}

Using the adopted value of $T_{e}$ (7000~K) we can evaluate the $T_{b}$ from
the RRL line integral at each pixel in the data cube.  Fig. \ref{fig7} shows a
contour plot of $T_{b}$ over the area of the data cube ($\ell =
36\degr$ to $44\degr$, $b = -4\degr$ to $4\degr$).  In addition to the compact sources, the diffuse emission can be seen peaking at $b \sim
0\degr$. Fig. \ref{fig8} plots the total continuum emission for this region, from this survey, as discussed in section \ref{sec:pkscont}. The similarities between the two maps show that the RRLs successfully recover both individual \hii~regions and the diffuse emission.

The extended object at $\ell \sim 40\degr$, $b=-2\degr$, W50, appears to have no RRL emission associated with it, so no evidence for ionized gas. This SNR is thought to be powered by the X-ray binary system SS433, which generates two relativistic jets in opposite directions, creating the lobes that elongate the feature. W50 lies at a distance of 6 kpc from the Sun and thus has no association with the \hii~region S74, which is seen in the RRL map at $(\ell,b)=(40\fdg0,-1\fdg3)$ \citep*{2007MNRAS.381..881L}. S74 is comprised of a compact source embedded in an extended region, also denominated RCW182. We measure a velocity of V$_{LSR}=44.7$\kms~for S74, which places it at a distance of $\sim 3.0$ kpc from the Sun, in the Sagittarius arm.

The region labelled G40.0-3.2 in Table 4 has sometimes been proposed as a possible \hii~region. \citet{2009BASI...37...45G} suggests that this feature lying to the SE of W50 is an SNR. The present RRL data shows no significant excess emission, although it lies on a weak diffuse free-free background. A $2\sigma$ upper limit to the RRL emission leads to $T_{b} \leq 0.3$ K compared with a total of 3 K. The \citet*{1986MNRAS.218..393D} conclusion based on the detection of 10 - 20 \% linear polarization and flux density spectral index of -0.6 ($S(\nu) \propto \nu^{-\alpha}$) is unaltered by this small free-free contribution. The 1.4 GHz continuum from this survey taken with the 2.7 GHz data of \citet{1984A&AS...58..197R} give a brightness temperature spectral index of $-3.0 \pm 0.2$ including a fitting error and assuming a 10 per cent calibration error at each frequency, thereby confirming the SNR identification.

\subsubsection{$T_{b}$ latitude distribution}
\label{sec:tb_lat_dist}
\begin{figure}
\centering
\includegraphics[scale=0.58]{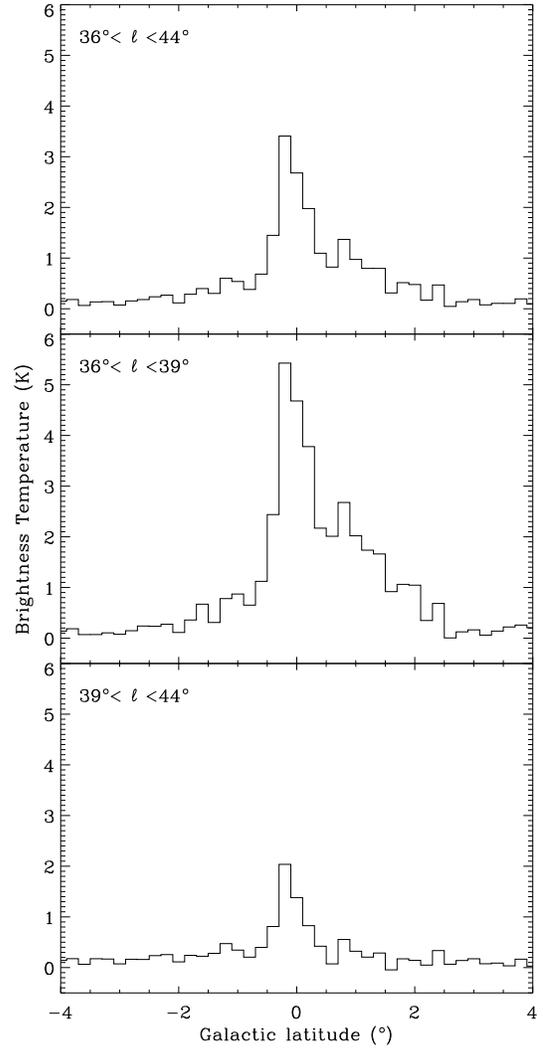}
\caption{Brightness temperature estimated from the ZOA RRLs versus latitude, averaged over three longitude ranges.}
\label{fig9}
\end{figure}

\begin{figure}
\centering
\includegraphics[scale=0.58]{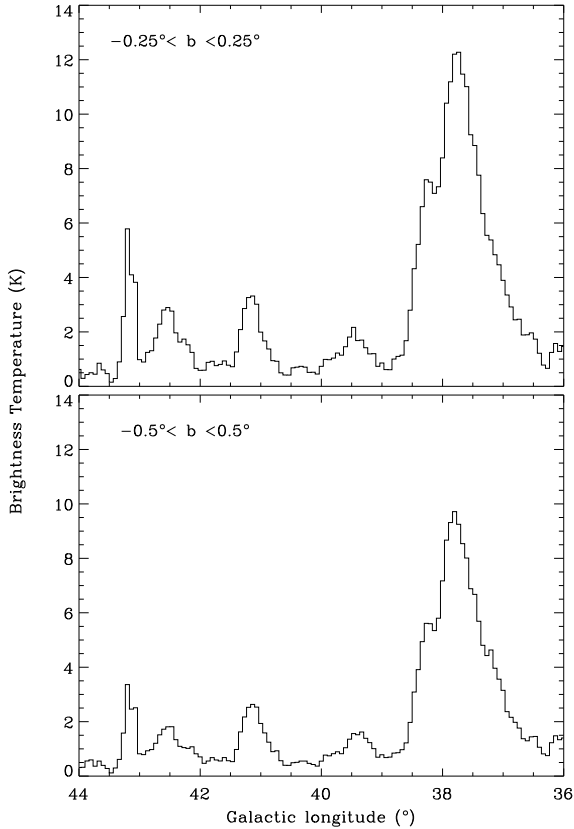}
\caption{Brightness temperature estimated from the ZOA RRLs versus longitude, averaged over two latitude ranges.}
\label{fig10}
\end{figure}

In order to make the diffuse emission clearer, latitude cuts
integrating over a longitude range are given in Fig. \ref{fig9}.  Fig. \ref{fig9}(a) is
an integral over the full longitude range $\ell = 36\degr$ to
$44\degr$ using $0\fdg2$ latitude bands.  Fig. \ref{fig9}(b) shows stronger $T_{b}$ in the $\ell$-range $36\degr$ to $39\degr$ than for $\ell
= 39\degr$ to $44\degr$ (Fig. \ref{fig9}(c)).  It should be noted that the
stronger emission at $\ell = 36\degr$ to $39\degr$ includes the
tangent to the Scutum arm which does not extend to $\ell = 39\degr$ to
$44\degr$.  The longitude variation of $T_{b}$ is shown in Fig. \ref{fig10} where (a) shows the Galactic ridge emission integrated from $b =
-0\fdg25$ to $0\fdg25$ while (b) is the integral from $b =
-0\fdg5$ to $0\fdg5$;  in both plots the longitude step is
4 arcmin.  We see the fall of $T_{b}$ with longitude as the line of sight
passes out of the Scutum arm, leaving only emission from the Sagittarius
and Local arms.

\begin{table*}
\centering
 \caption{\hii~regions and Supernova Remnants in the region $\ell=36\degr$ to $44\degr$, from the \citet{1970A&AS....1..319A}, \citet{2009BASI...37...45G} and \citet{2003A&A...397..213P} catalogs. The table arrangement is as follows: columns 1 and 2 - source coordinate in Galactic longitude and latitude; column 3 - source angular diameter (in declination and right ascension), where the letter p stands for point source; column 4 - source integrated flux density at 1.4 GHz (sources with integrated flux density greater than 1 Jy only); column 5 - measured LSR velocity from a gaussian fit to the RRL (when two values are presented it means that two velocity components are detected in the RRL spectrum); column 6 - published LSR velocity (where available, from the \citet{2003A&A...397..213P} synthetic catalog); column 7 - distance from the Sun (calculated as in \citet{2004MNRAS.347..237P} for the \hii~regions for which the distance ambiguity is resolved by auxiliary data) and from the \citet{2009BASI...37...45G} catalog for the SNRs, where available); column 8 - source identification. 
\label{tab:HIIreg}}
  \begin{tabular}{cccccccc}
    \hline
 $\ell$ & $b$ & Diameter & Integrated flux & Measured LSR & Published LSR & Distance & Notes  \\
 (\degr) & (\degr) & (arcmin) & density (Jy) & velocity (km/s) & velocity (km/s) & (kpc) &         \\   \hline

 36.3 & -1.7 & 15 \ 16  & 11  & 63.7 &$63.5$  & 4.1 &\hii~RCW179\\
 36.3 & +0.7 & 8 \ 8    & 3   & 71.3 &$76.5$  &  &\hii\\
 36.6 & -0.7 & 12 \ 17  & 5   & - &  -           &  &SNR\\
 37.0 & -0.2 & p \ 14   & 4   & 36.6, 82.6 &              &  &\hii\\
 37.5 & -0.1 & 6 \ 10   & 12  & 52.9 &$49.0$  & 10.3 &\hii~W47\\
 37.5 & +0.9 & 10 \ 22  & 4   & 44.2 &              &  &\hii\\
 37.7 & +0.1 & 6 \ 7    & 6   & 44.1, 87.7 &$88.9$  &  &\hii~W47\\
 37.8 & -0.3 & 5 \ 17   & 18  & 60.8 &              &  &\hii~W47\\
 38.1 & -0.1 & 17 \ 14  & 3   & 58.6 &              &  &\hii\\
 38.4 & +1.3 & 10 \ 10  & 2   & 42.1 &              &  &\hii\\
 39.2 & -0.3 & 8 \ 6    & 15  & - &  -            & $>$ 7.7 &SNR\\
 39.3 & +0.0 & 2 \ 2    & 3   & 16.8 &              &  &\hii\\
 39.5 & +0.5 & 11 \ 14  & 4   & 45.4 &              &  &\hii\\
 39.7 & -2.0 & 120 \ 60 & 67  & - & -             & $6.0$ &SNR W50\\
 40.0 & -3.2 & 23 \ 23  & 10  & &              &  &\hii~(?)\\
 40.0 & -1.3 & 21 \ 17  & 9   & 44.7 &  44.3         &  3.0 &\hii~S74, RCW182\\
 40.5 & -0.5 & 22 \ 22  & 9   & - & -            &  &SNR\\
 40.5 & +2.5 & 9 \ 10   & 10  & 23.4 &$23.0$  & 1.6 &\hii~W45\\
 40.6 & -0.5 & 10 \ 15  & 8   & 86.0 &             &  &\hii\\
 41.1 & -0.2 & 3.7 \ 3.7& 5.3 & 56.4 &  59.4  & 8.8   &\hii\\
 41.1 & -0.3 & 4.5 \ 2.5& 19  & - & -            & $>$ 7.5 &SNR\\
 41.4 & +0.4 & 17 \ 12  & 9   & 28.4, 81.7 &              &  &\hii\\
 42.0 & -0.1 & 25 \ 18  & 11  & 32.4 &              &  &\hii\\
 42.1 & -0.6 & 6 \ 5    & 6   & 67.5 &$66.0$  & 4.7 &\hii\\
 42.2 & +0.0 & p \ p    & 2   & 49.8 &             &  &\hii\\
 42.5 & -0.2 & 19 \ 12  & 7   & 22.8, 68.7 &             &  &\hii\\
 42.8 & +0.6 & 24 \ 24  & 2.5 & - & -              &  &SNR\\
 43.2 & -0.5 & p \ p    & 2   & 56.3 &$58.0$  &  &\hii\\
 43.2 & +0.0 & 2 \ 2    & 47  & 6.5 &$7.7$   & 11.9 &\hii~W49A\\
 43.3 & -0.2 & 4 \ 3    & 32  & - & -            & 10 &SNR W49B\\
 43.9 & +1.6 & 60 \ 60  & 8 & - & -            &  &SNR\\

\hline
\end{tabular}
\end{table*}

\subsection{Signal loss in the ZOA}
\label{sec:signalloss}

The strongest emission seen in the $\ell$-range $36\degr$ to $39\degr$ compared to $\ell
= 39\degr$ to $44\degr$ is intrinsically due to the presence of the Scutum arm.
Nonetheless, the emission in $39\degr < \ell <44\degr$ appears to be low, with the peak of the latitude distribution less than half of that in the range $36\degr < \ell <39\degr$. There is also less extended emission around the plane, so the latitude distribution appears narrow and is mainly accounted for by the few \hii~regions in that longitude range.

This suggests that we might be losing some extended diffuse emission in the process, probably due to the bandpass removal associated with the scanning strategy.
As described in section \ref{sec:data_red}, the adopted bandpass removal technique, {\sc minmed5}, is appropriate for recovering extended emission, but only when it does not fill the entire scan. This is likely to be the cause of some flux  density loss seen on the Galactic plane in the ZOA and the presence of negative sidelobes. We discuss this further in Section \ref{sec:wmapMEM}. Also, taking the minimum of a series of medians as the bandpass correction for an entire scan can introduce small variations between adjacent scans (striping).

\section{The latitude distribution of $T_{b}$ - comparison with other data}
\label{sec:latitude_distribution}

In the previous sections we have derived the distribution of RRL line
integral on and near the Galactic plane at $\ell = 36\degr$ to
$44\degr$.  This gives an unambiguous measure of $T_{b}$ when scaled for
electron temperature.  
Since this RRL emission is weak we will average it in longitude to provide a more sensitive determination of the latitude distribution up to $|b| = 4\degr$.
In this section other data are used to compare and extend our analysis of the free-free emission, namely \ha, WMAP Maximum Entropy Model and the 21 cm continuum.

\subsection{The \ha~latitude distribution}
\label{sec:ha_latitude_distribution}

We follow the approach of \citet{2003MNRAS.341..369D} to derive the
\ha~latitude distribution.  WHAM, the sensitive \ha~survey of the
northern sky provides a map with accurately determined baselines at a
resolution of $\sim$ 60~arcmin \citep{2003ApJS..149..405H}.  This map is then
corrected for absorption by Galactic dust using the \citet{1998ApJ...500..525S} $100~\mu$m map corrected to a fixed dust
temperature of 18.2~K. The total absorption (for an extragalactic
object) at the wavelength of \ha~is estimated as $(0.0462 \pm 0.0035)
D^{\rm T}$ magnitudes where $D^{\rm T}$ is the $100~\mu$m temperature-corrected intensity in
MJy~sr$^{-1}$.  Dickinson et al. find that the effective absorption is
$\sim 0.3$ of this value. This 0.3 factor, $f_{d}$, represents the relative distribution of the gas (\ha) and dust in the line of sight at intermediate and high Galactic latitudes. When the effective absorption
reaches 1 magnitude, the correction becomes uncertain and the estimate
of the true \ha~intensity is unreliable.

The corrected \ha~intensity
$I_{\rm H\alpha}$ is converted to $T_{b}$ using the formula given by
Dickinson et al., namely

\begin{equation}
T_{b} = 8.396 \times 10^{3} I_{\rm H_{\alpha}} a \nu_{GHz}^{-2.1} T_{4}^{0.667} 10^{0.029/T_{4}} (1+0.08)
\label{eq:8}
\end{equation}

where $T_{4}$ is the electron temperature in units of $10^{4}$~K and
$I_{{\rm H}\alpha}$ in units of erg~cm$^{-2}$~$s^{-1}$~sr$^{-1}$ (or Rayleigh, R, 1 R $\equiv 2.41 \times 10^{7}$ erg~cm$^{-2}$~$s^{-1}$~sr$^{-1}$). At 1.4 GHz and $T_{e} = 7000$ K, the values used here, the free-free brightness temperature per unit Rayleigh is 3.78 mK.

\begin{figure}
\centering
\includegraphics[scale=0.58]{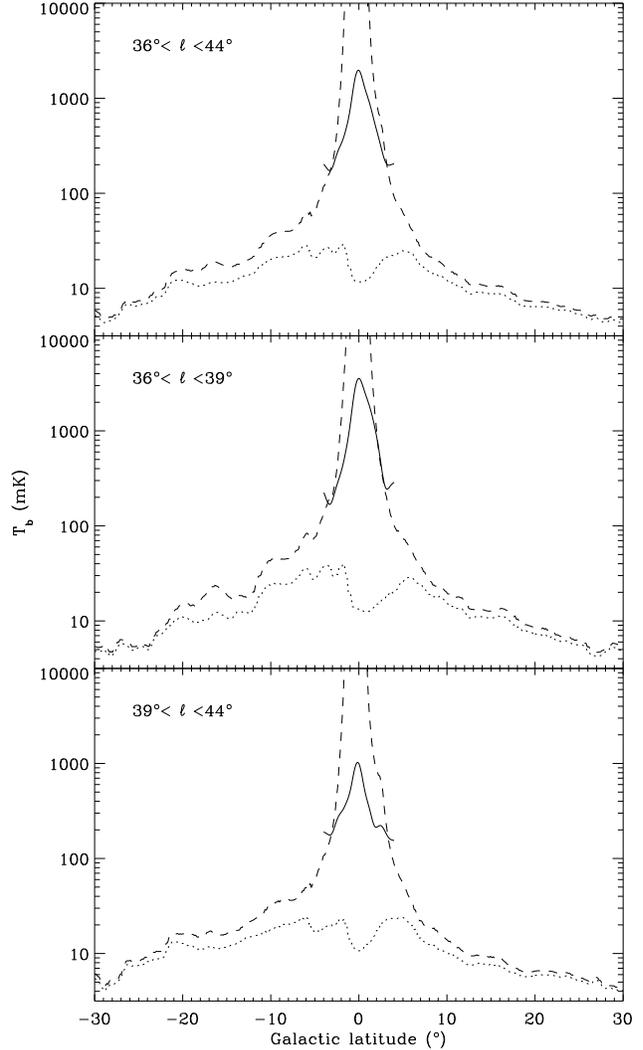}
\caption{The \ha~latitude distribution, uncorrected (dotted line) and corrected for dust absorption (dashed line) as well as the free-free estimates derived from the RRLs (solid line), averaged for three longitude ranges. The WHAM intensity is corrected for dust absorption and converted to brightness temperature for $T_{e} = 7000$ K, $\nu = 1.4$ GHz and $f_{d}$ values of 0.73 and 0.47 for the negative and positive sides of the distribution, respectively. Note the logarithmic vertical scale.}
\label{fig11}
\end{figure}

Fig. \ref{fig11} shows the latitude distribution of $T_{b}$, between $b = -30\degr$ to
$30\degr$ averaged over the longitude ranges $\ell = 36\degr$ to
$44\degr$, $36\degr$ to $39\degr$ and $39\degr$ to $44\degr$, for the uncorrected \ha~(dotted line), the corrected \ha~(dashed line) and the RRLs free-free estimation (solid line). 
The angular resolution of Fig. \ref{fig11} is 60~arcmin.  
We use the free-free estimates derived from the RRLs, smoothed to 1\degr, and the \ha~data, to derive the $f_{d}$ value that best fits the two distributions between latitudes of $|b| = 3\degr$ to $4\degr$. The best fit $f_{d}$ values for the negative and positive sides of the latitude distribution are $0.73^{+0.11}_{-0.56}$ and $0.47^{+0.19}_{-0.35}$, respectively. The \ha~data in Fig. \ref{fig11} are corrected for dust absorption using the derived $f_{d}$ values. The effective absorption is about 1 magnitude at $|b| \simeq 5\degr$ and increases to 2 magnitudes at $|b| \simeq 4\degr$. At this latitude, the brightness temperature from the \ha~and RRLs is $\sim 200$ mK.

A uniform mixing of gas and dust corresponds to an $f_{d}$ value of 0.5. Our derived values at $|b| = 3\degr$ to $4\degr$ are similar to uniform mixing. However, at lower latitudes the values determined from the \ha~data are far higher than the observed free-free emission from RRLs (and WMAP MEM free-free). This would appear to indicate that the FDS dust model for $D^{\rm T}$ overestimates the absorption produced by the warm (40 K) dust on the Galactic ridge. This is not unexpected since the dust properties such as temperature, composition, ISM radiation field, along this deep line of sight through the Galaxy are different from those at intermediate latitudes covered by the FDS model. 

\begin{figure}
\centering
\includegraphics[scale=0.5]{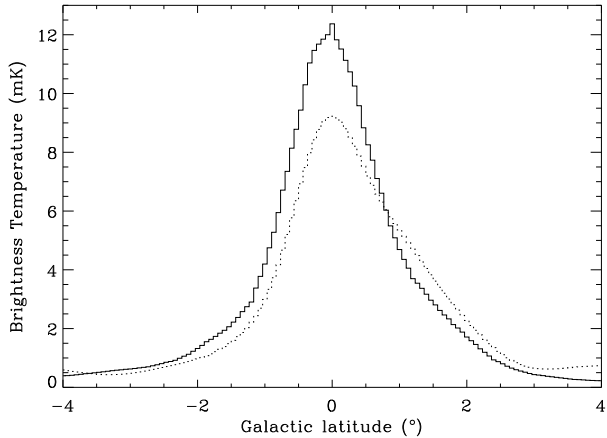}
\caption{Free-free emission estimated from the RRLs and from the WMAP 23 GHz Maximum Entropy Model versus latitude, averaged over $36\degr$ to $39\degr$ longitude. The full line represents the WMAP MEM and the dotted line is the free-free emission from the ZOA smoothed to 1\degr resolution and an assumed $T_{e} = 7000$ K.}
\label{fig12}
\end{figure}

\begin{figure}
\centering
\includegraphics[scale=0.58]{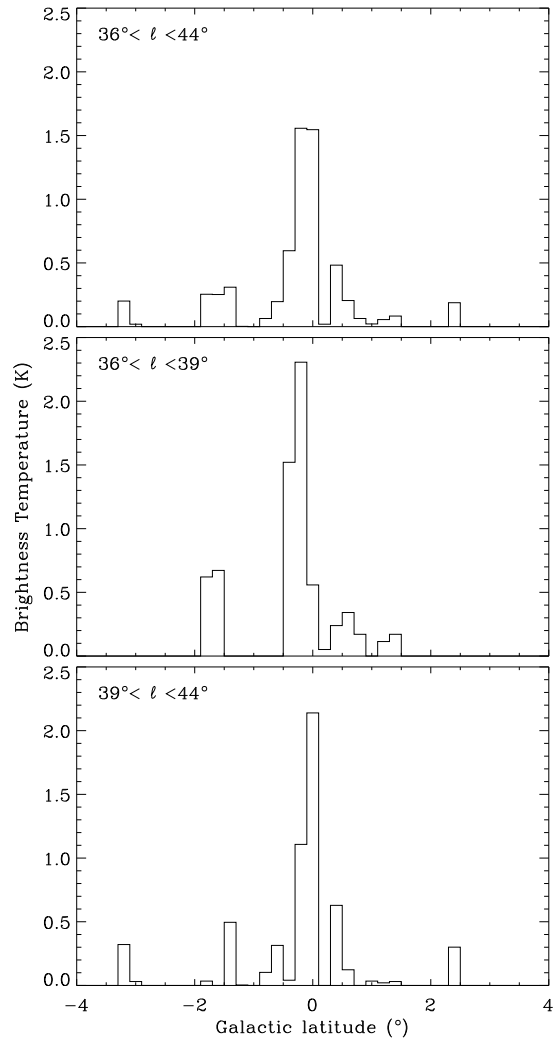}
\caption{The contribution from discrete \hii~regions to the diffuse ISM free-free emission at 1.4 GHz. Brightness temperatures are calculated for 84 \hii~regions from the 2.7 GHz catalog of \citet{2003A&A...397..213P} and plotted against Galactic latitude, using $0\fdg2$ latitude bands.}
\label{fig13}
\end{figure}

\subsection{WMAP Free-free MEM Model}
\label{sec:wmapMEM}

The WMAP team has estimated the contribution of synchrotron, thermal dust and free-free emission, at each band and at 1\degr~resolution, using a Maximum Entropy Method (MEM) \citep{2009ApJS..180..265G}. The MEM is an iterative procedure that fits for the emission of each component and also the synchrotron spectral index, assuming that the dust and free-free temperature follows a power law with $\beta=+2$ and $\beta=-2.14$, respectively.
Where the WMAP's S/N is low, templates made from external data are used as priors. These templates are: the Haslam 408 MHz map \citep{1982A&AS...47....1H} for the synchrotron, the extinction-corrected \ha~map \citep{2003ApJS..146..407F} for the free-free and the Model 8 of \citet*{1999ApJ...524..867F} map for the thermal dust.

Fig. \ref{fig12} shows, for the longitude range $36\degr$ to $39\degr$, the brightness temperature estimation from the ZOA RRLs (dotted line) with an assumed spectral index $\beta=-2.13$ and WMAP's MEM (full line), at 23 GHz. 
The brighter longitude range $36\degr$ to $39\degr$ is chosen because of the lower signal loss in this range (see Section \ref{sec:signalloss}).
We can see that the peak, at $b=0\degr$, of the MEM distribution is about 30 per cent higher than that of the ZOA. It is essential to point out that the $T_{b}$ estimation from the RRLs depends on $T_{e}^{1.15}$ and that the electron temperature may vary across the $8\degr$ region. As mentioned in Section \ref{sec:em}, it is likely to be higher for the diffuse emission than for individual \hii~regions. If we increase the electron temperature to 8000 K, the ZOA distribution will then increase by a factor of 1.16, which makes the integral of the two distributions agree within $10\%$.
Another difference between the two distributions is the fact that the ZOA is not as symmetric around the plane as the MEM is. This is due, as seen before, to the feature at $\ell \sim 38\degr$ that extendeds from the plane up to $b \sim 2\degr$. As a result, the FWHM of the ZOA latitude distribution for this longitude range is $\sim 2\degr$, whereas the MEM has a FWHM of $\sim 1\fdg7$. With the lack of information about free-free emission on the plane, it is not surprising if the Maximum Entropy Method cannot separate synchrotron and free-free accurately. Moreover, no error estimates are available for the MEM maps. Nevertheless, there is relatively good agreement between the two maps.

\subsection{The contribution of individual \hii~regions}
\label{sec:hii_regions}

Our analysis so far has included the individual \hii~regions that
occur within the survey area.  The extensive catalogue of \citet{2003A&A...397..213P} 
can be used to derive the latitude distribution of
free-free emission in the RRL data cube.  Flux densities  at 2.7~GHz
are given for 84 \hii~regions in the catalog for the longitude range
$36\degr$ to $44\degr$.  These can be used to generate an average
2.7~GHz brightness temperature profile in latitude using the expression
\begin{equation}
S  =  \frac{2k T_{\rm b} d\Omega}{\lambda^{2}} \label{eq:10}
\end{equation}                                             
This is converted into a $T_{b}$ latitude profile at 1.4 GHz, extrapolated using a spectral index $\beta=-2.10$, for 3 longitude ranges
($\ell = 36\degr$ to $39\degr$, $39\degr$ to $44\degr$, $36\degr$ to
$44\degr$) in Fig. \ref{fig13}.  It is seen that the \hii~region contribution
amounts to $20-30$ percent of the total free-free ($T_{b}$) emission in
this longitude range.

\subsection{Radio continuum at 1.4 GHz}
\label{sec:pkscont}

The 1.4 GHz continuum map for the area $\ell = 36\degr$ to $44\degr$, $b = -4\degr$ to $4\degr$, shown in Fig. \ref{fig8}, is produced by combining the ZOA and HIPASS data for this region.

\begin{figure}
\includegraphics[scale=0.5]{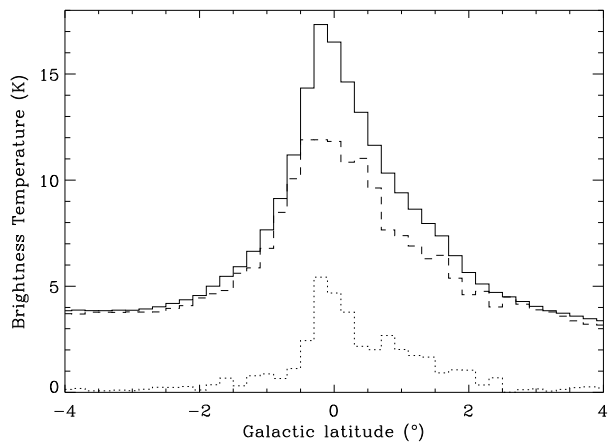}
\caption{Comparison between the continuum and the free-free estimation from the ZOA RRLs, versus latitude, for the longitude range $\ell = 36\degr$ to $39\degr$, using $0\fdg2$ latitude bands. The full line represents the continuum, the dotted is the RRL free-free estimate and the dashed is the difference between the two, representing the synchrotron emission. }
\label{fig14}
\end{figure}

We are now in a position to separate the 1.4 GHz free-free and synchrotron emission components in the present survey area. The synchrotron emission is the total intensity minus the 1.4 GHz free-free estimated from the RRL data. Fig. \ref{fig14} shows the latitude distribution of $0\fdg2$ bins for the $\ell$ range $36\degr$ to $39\degr$. We find that the derived synchrotron emission is the major contributor at 1.4 GHz, accounting for $\sim 70\%$ of the total emission, on the plane. At higher latitudes, where the free-free emission has fallen, it is dominant. Its latitude distribution appears to have two components - one with a FWHP of $\sim 10\degr$ and a narrow component confined to the plane with a FWHP $\sim 2\degr$ when corrected for the contribution of SNRs in this longitude range.

\begin{figure*}
\centering
\hspace{1.0cm}
\includegraphics[scale=0.8]{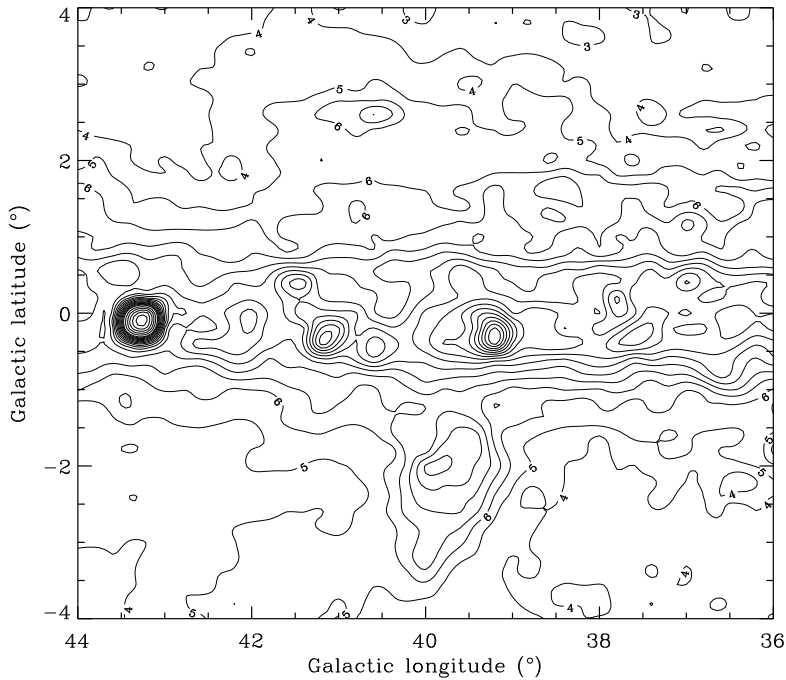}
\caption{Map of the 1.4 GHz synchrotron emission in the survey region; it is the continuum emission minus the RRL free-free estimate. Contours are as in Fig. \ref{fig8}; the first four contours are labelled. The resolution is 15.5 arcmin.}
\label{fig15}
\end{figure*}

Fig. \ref{fig15} shows a map of the derived 1.4 GHz synchrotron emission at a resolution of 15.5 arcmin. The narrow diffuse emission confined to the plane is evident. In addition, a number of known SNRs (Table \ref{tab:HIIreg} and \citealt{2009BASI...37...45G}) can be identified. Additional possible SNRs identified by \citet{2002ApJ...566..378K} (($\ell$,$b$) = ($41\fdg5$,$+0\fdg4$), ($42\fdg0$,$-0\fdg1$) and ($43\fdg5$,$+0\fdg6$)) can be seen; these and other candidates will be discussed in a future paper.

\subsection{WMAP: 23 to 94 GHz - Identification of the 4 foreground components}
\label{sec:wmap}

Our determination of the free-free emission from the RRL data makes a significant contribution to the derivation of the Galactic foregrounds on the Galactic plane at higher frequencies. The WMAP 5-yr \citep{2009ApJS..180..225H} data at 23, 33, 41, 61 and 94 GHz have the resolutions of $\sim$ 49, 37, 29, 20 and 13 arcmin respectively, which are adequate to investigate the Galactic latitude distribution which has a width of $\sim 2\degr$ in the longitude range of the present study. We use the 1\degr resolution WMAP data for each of the five frequency bands\footnote{Downloaded from the website http://lambda.gsfc.nasa.gov.}. 

The synchrotron component is given by the 408 MHz survey \citep{1982A&AS...47....1H}, made with a resolution of 51 arcmin; its brightness temperature spectral index is assumed to be $\beta = -3.0$ (\citealt{2003MNRAS.345..897B,2006MNRAS.370.1125D}). The free-free emission is that derived from our RRL study with a resolution of 15.5 arcmin. Its spectral index is assumed to be $\beta = -2.13$ over the frequency range of WMAP \citep{2003MNRAS.341..369D}. Both the synchrotron and the free-free templates are smoothed to 1\degr~for this comparison.

\begin{figure}
\centering
\includegraphics[scale=0.72]{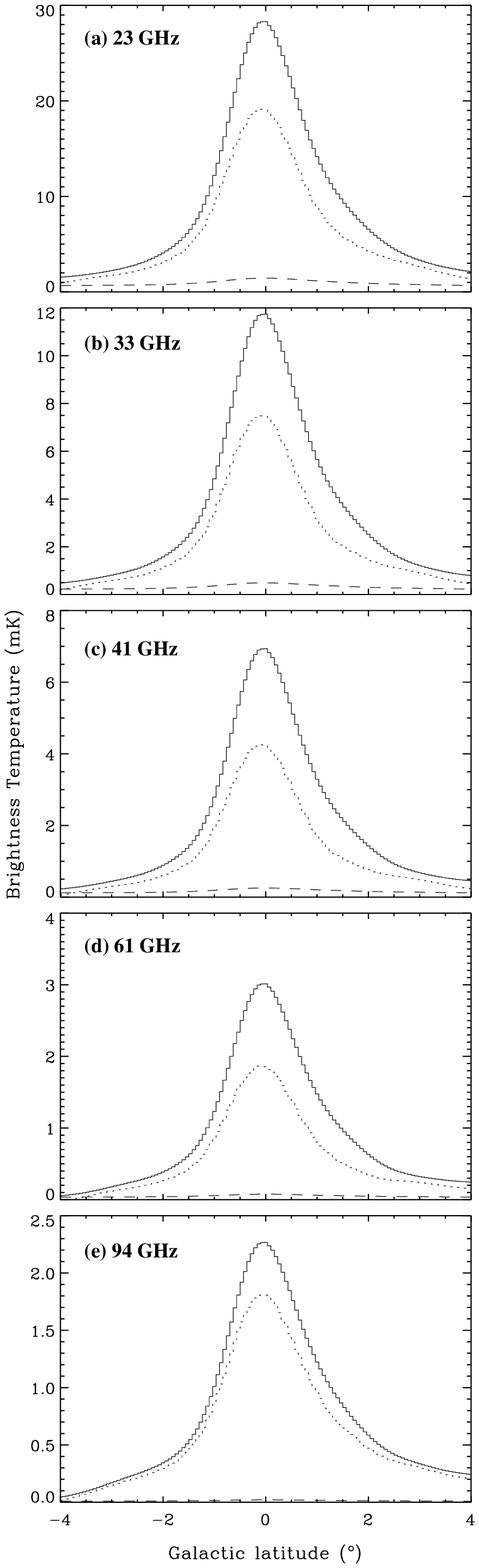}
\caption{Comparison between WMAP, synchrotron and WMAP RRL-corrected versus latitude, at each frequency (23 GHz (a) to 94 GHz (e)), averaged over the longitude range $\ell$-range $36\degr$ to $39\degr$, at 1\degr~resolution. The full line represents the WMAP temperature (I) 5-year data; the dotted line is the result of WMAP minus the the free-free; the dashed line is the 408 MHz data scaled with a spectral index $\beta = -3.0$.}
\label{fig16}
\end{figure}

Figs. \ref{fig16}(a) to (e) show the latitude cuts at each WMAP frequency for the longitude range 36\degr to 39\degr. The full line gives the total WMAP brightness temperature, while the dotted line gives the total minus the free-free and the dashed line gives the synchrotron emission, at each frequency. It is seen immediately that the synchrotron is a minor contributor at WMAP frequencies - amounting to only 5 per cent at 23 GHz and correspondingly less at higher frequencies falling to only 1 per cent at 94 GHz. The dotted line minus the synchrotron (dashed line), represents the excess "anomalous" emission. 

The origin of the anomalous emission is not entirely clear. A strong candidate is spinning dust (\citealt*{1998ApJ...494L..19D,2009MNRAS.395.1055A}) which is expected to peak in the frequency range 20-40 GHz and to be correlated with FIR dust emission. At 23, 33 and 41 GHz this anomalous emission is 65, 59 and 57 per cent, respectively, of the total emission. Expressed more physically, the anomalous emission at these three frequencies is 2.2, 1.6 and 1.2 of the free-free emission. Assumptions on the free-free spectral index, electron temperature and calibration errors do not affect the results on the anomalous emission detection by more than 20 per cent.

At 61 and 94 GHz the thermal (vibrational) dust emission is a significant contributor. On the plane this dust is warmer ($T_{d} \simeq 40$ K) than at intermediate latitudes (20 K) having been heated by the radiation field responsible for ionizing the ISM to produce the free-free. At 94 GHz the excess emission is 4.2 times the free-free. If a thermal dust emissivity spectral index of $\beta = +2.0$ is assumed, it will contribute 45 per cent of the observed excess at 61 GHz. On the other hand, if $\beta = +1.5$, it contributes 65 per cent and will have a significant contribution at 41 GHz. 

\begin{figure}
\centering
\includegraphics[scale=0.70]{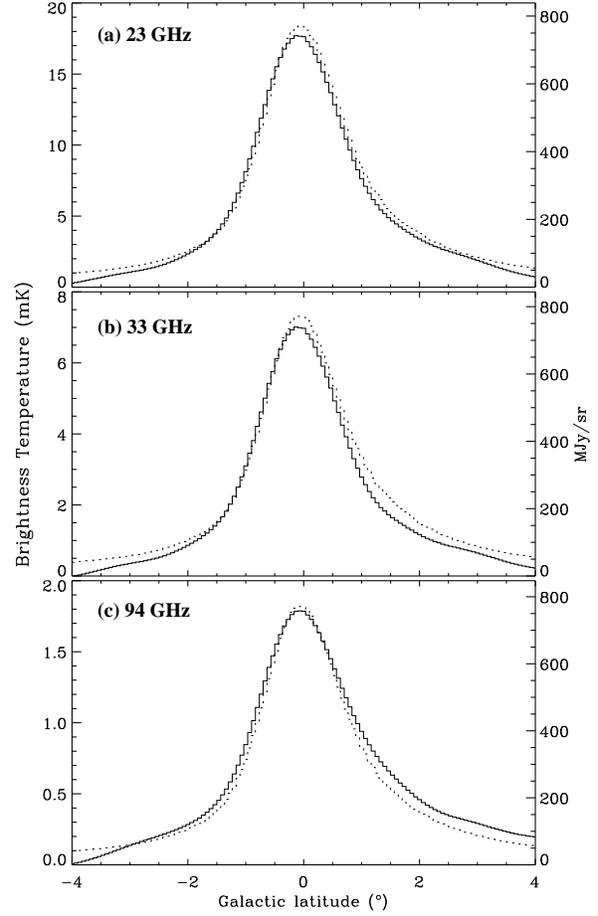}
\caption{The anomalous emission (i.e., total minus free-free minus synchrotron) latitude distribution for the longitude range 36\degr~to 39\degr~at 23 GHz (a), 33 GHz (b) and 94 (c) GHz (full line), compared with the 100 $\mu$m surface brightness distribution (dotted line).}
\label{fig17}
\end{figure}

A critical test for the spinning dust model is that the width of the observed anomalous radio emission should be similar to that of the (warm) dust. The 100 $\mu$m IRIS/IRAS data \citep{2005ApJS..157..302M}, again smoothed to 1\degr, are compared with the WMAP data in Figs. \ref{fig17}(a), (b) and (c) at 23, 33 and 94 GHz, for the longitude range 36\degr~to 39\degr. The latitude distributions of the radio (full line) and dust emissions (dotted line) are clearly similar both at 23 and 33 GHz (spinning dust) and at 94 GHz (thermal dust).
The half power widths are $1\fdg78$, $1\fdg83$, $1\fdg77$ and $2\fdg01$ for 100$\mu$m, 23 GHz, 33 GHz and 94 GHz respectively. The ratio of 33 GHz to 100 $\mu$m brightness is 9.5 $\mu$K(MJy sr$^{-1}$)$^{-1}$, a value similar to that found at intermediate latitudes \citep{2006MNRAS.370.1125D}.
These WMAP data are in agreement with a spinning dust model.

\section{Discussion and conclusions}
\label{sec:discussion_conclude}

The Parkes HIPASS and Zone of Avoidance survey of HI contains 3 RRLs which can 
be summed to provide a map of the RRL emission over the ZOA region at an angular 
resolution of 15.5 arcmin. The $8\degr \times 8\degr$ data cube centred on 
$(\ell,b)=(40\degr, 0\degr)$ has been used to illustrate the viability of such an investigation. 
This work has demonstrated that the 1.4 GHz ZOA survey is of sufficient sensitivity to be 
able to provide a map of free-free emission on the Galactic plane in the inner ($\ell < 50\degr$) 
Galaxy. The velocity resolution is shown to be sufficient to be able to identify the emission from the 
Local, Sagittarius and Scutum spiral arms.

In order to convert this RRL line integral to an emission measure for comparison with other data, 
it is necessary to have a value of the electron temperature, $T_{e}$. This can be obtained from a 
comparison with free-free brightness temperature scaled to 1.4 GHz derived from the WMAP MEM 
free-free model for the same region of the Galactic plane. The best fit model gives $T_{e} = 8000$ K 
for the total free-free emission. However the $T_{e}$ for individual \hii~regions varies with 
Galactocentric radius such that the mean $T_{e}$ for the Local, Sagittarius and Scutum spiral arms is 
6600 K, 6000 K and 5500 K, respectively. Since these \hii~regions represent 20 - 30 percent of the 
total free-free emission, the remaining diffuse emission on the plane must have $T_{e}$ somewhat greater than 8000 K. 
Such higher values for the diffuse emission have been discussed in the literature \citep{1985ApJ...294..256R}.

Our clear determination of the free-free emission in the $(\ell,b)=(40\degr, 0\degr)$ data cube, 
allows us to investigate the anomalous emission in this section of the Galactic plane by using 
the WMAP data along with the 408 MHz synchrotron template. We have shown that excess (above the 
free-free and synchrotron) emission has the same latitude distribution as the 100$\mu$m FIR IRAS 
emission. This is expected for a dust-correlated component of emission. The lower WMAP frequencies 
(23 and 33 GHz) fit a spinning dust model for very small dust grains. The highest frequency (91 GHz) 
emission is the long wavelength tail of the thermal emission from large dust grains. The 61 GHz 
emission is composed of roughly equal amounts from both mechanisms.

The present study will now be extended to cover more of the deep ZOA $|b| < 4\degr$ 
region. It will be supplemented by the wider HIPASS survey which, although it has a 
quarter of the integration time of the ZOA, it nevertheless covers the whole southern sky 
and therefore can give an independent estimate of zero levels. In combination with the 
\ha~all-sky data which give a measure of the free-free emission when corrected for absorption, 
the RRL data cover the remaining low latitude regions, where the correction is uncertain, 
thereby providing a complete all-sky free-free emission template.


\section*{ACKNOWLEDGMENTS}

We thank Virginia Kilborn and Dave Barnes for help during the early stages of the project.
MIRA is funded by the Funda\c{c}\~{a}o para a Ci\^{e}ncia e Tecnologia (Portugal). CD acknowledges a STFC Adanced Fellowship. The Parkes telescope is part of the Australia Telescope which is funded by the Commonwealth of Australia for operation as a National Facility managed by CSIRO.



\begin{thebibliography}{}


\bibitem[\protect\citeauthoryear{{Ali-Ha{\"i}moud}, {Hirata} \&
  {Dickinson}}{{Ali-Ha{\"i}moud} et~al.}{2009}]{2009MNRAS.395.1055A}
{Ali-Ha{\"i}moud} Y.,  {Hirata} C.~M.,    {Dickinson} C.,  2009, \mnras, 395,
  1055

\bibitem[\protect\citeauthoryear{{Altenhoff}, {Downes}, {Goad}, {Maxwell} \&
  {Rinehart}}{{Altenhoff} et~al.}{1970}]{1970A&AS....1..319A}
{Altenhoff} W.~J.,  {Downes} D.,  {Goad} L.,  {Maxwell} A.,    {Rinehart} R.,
  1970, \aaps, 1, 319

\bibitem[\protect\citeauthoryear{{Banday}, {Dickinson}, {Davies}, {Davis} \&
  {G{\'o}rski}}{{Banday} et~al.}{2003}]{2003MNRAS.345..897B}
{Banday} A.~J.,  {Dickinson} C.,  {Davies} R.~D.,  {Davis} R.~J.,
  {G{\'o}rski} K.~M.,  2003, \mnras, 345, 897

\bibitem[Barnes et al.(2001)]{2001MNRAS.322..486B} Barnes, D.~G., et al.\ 
2001, \mnras, 322, 486 


\bibitem[\protect\citeauthoryear{{Bronfman}, {Casassus}, {May} \&
  {Nyman}}{{Bronfman} et~al.}{2000}]{2000A&A...358..521B}
{Bronfman} L.,  {Casassus} S.,  {May} J.,    {Nyman} L.-{\AA}.,  2000, \aap,
  358, 521

\bibitem[\protect\citeauthoryear{{Davies}, {Dickinson}, {Banday}, {Jaffe},
  {G{\'o}rski} \& {Davis}}{{Davies} et~al.}{2006}]{2006MNRAS.370.1125D}
{Davies} R.~D.,  {Dickinson} C.,  {Banday} A.~J.,  {Jaffe} T.~R.,  {G{\'o}rski}
  K.~M.,    {Davis} R.~J.,  2006, \mnras, 370, 1125

\bibitem[\protect\citeauthoryear{{Dennison}, {Simonetti} \&
  {Topasna}}{{Dennison} et~al.}{1998}]{1998PASA...15..147D}
{Dennison} B.,  {Simonetti} J.~H.,    {Topasna} G.~A.,  1998, Publications of
  the Astronomical Society of Australia, 15, 147

\bibitem[\protect\citeauthoryear{{Dickinson}, {Davies} \& {Davis}}{{Dickinson}
  et~al.}{2003}]{2003MNRAS.341..369D}
{Dickinson} C.,  {Davies} R.~D.,    {Davis} R.~J.,  2003, \mnras, 341, 369

\bibitem[Donley et al.(2005)]{2005AJ....129..220D} Donley, J.~L., et al.\ 
2005, \aj, 129, 220 


\bibitem[\protect\citeauthoryear{{Downes}, {Pauls} \& {Salter}}{{Downes}
  et~al.}{1986}]{1986MNRAS.218..393D}
{Downes} A.~J.~B.,  {Pauls} T.,    {Salter} C.~J.,  1986, \mnras, 218, 393

\bibitem[\protect\citeauthoryear{{Draine} \& {Lazarian}}{{Draine} \&
  {Lazarian}}{1998}]{1998ApJ...494L..19D}
{Draine} B.~T.,  {Lazarian} A.,  1998, \apjl, 494, L19+

\bibitem[\protect\citeauthoryear{{Finkbeiner}}{{Finkbeiner}}{2003}]{2003ApJS..%
146..407F}
{Finkbeiner} D.~P.,  2003, \apjs, 146, 407

\bibitem[\protect\citeauthoryear{{Finkbeiner}, {Davis} \&
  {Schlegel}}{{Finkbeiner} et~al.}{1999}]{1999ApJ...524..867F}
{Finkbeiner} D.~P.,  {Davis} M.,    {Schlegel} D.~J.,  1999, \apj, 524, 867 

\bibitem[\protect\citeauthoryear{{Gaensler}, {Madsen}, {Chatterjee}, \& {Mao}}{{Gaensler} et~al)}{2008}]{2008PASA...25..184G} {Gaensler}, B.~M., 
{Madsen}, G.~J., {Chatterjee} S., 
\& {Mao}, S.~A., 2008, Publications of the Astronomical Society of Australia, 25, 184 


\bibitem[\protect\citeauthoryear{{Gaustad}, {McCullough}, {Rosing} \& {Van
  Buren}}{{Gaustad} et~al.}{2001}]{2001PASP..113.1326G}
{Gaustad} J.~E.,  {McCullough} P.~R.,  {Rosing} W.,    {Van Buren} D.,  2001,
  \pasp, 113, 1326

\bibitem[Gold et al.(2009)]{2009ApJS..180..265G} Gold, B., et al.\ 2009, 
\apjs, 180, 265 

\bibitem[\protect\citeauthoryear{{Gordon} \& {Cato}}{{Gordon} \&
  {Cato}}{1972}]{1972ApJ...176..587G}
{Gordon} M.~A.,  {Cato} T.,  1972, \apj, 176, 587

\bibitem[\protect\citeauthoryear{{Green}}{{Green}}{2009}]{2009BASI...37...45G}
{Green} D.~A.,  2009, 37, 45

\bibitem[\protect\citeauthoryear{{Haffner}, {Reynolds}, {Tufte}, {Madsen},
  {Jaehnig} \& {Percival}}{{Haffner} et~al.}{2003}]{2003ApJS..149..405H}
{Haffner} L.~M.,  {Reynolds} R.~J.,  {Tufte} S.~L.,  {Madsen} G.~J.,  {Jaehnig}
  K.~P.,    {Percival} J.~W.,  2003, \apjs, 149, 405

\bibitem[\protect\citeauthoryear{{Hart} \& {Pedlar}}{{Hart} \&
  {Pedlar}}{1976}]{1976MNRAS.176..547H}
{Hart} L.,  {Pedlar} A.,  1976, \mnras, 176, 547

\bibitem[\protect\citeauthoryear{{Haslam}, {Salter}, {Stoffel} \&
  {Wilson}}{{Haslam} et~al.}{1982}]{1982A&AS...47....1H}
{Haslam} C.~G.~T.,  {Salter} C.~J.,  {Stoffel} H.,    {Wilson} W.~E.,  1982,
  \aaps, 47, 1

\bibitem[Hinshaw et al.(2009)]{2009ApJS..180..225H} Hinshaw, G., et al.\ 
2009, \apjs, 180, 225 


\bibitem[Kaplan et al.(2002)]{2002ApJ...566..378K} Kaplan, D.~L., Kulkarni, 
S.~R., Frail, D.~A., \& van Kerkwijk, M.~H.\ 2002, \apj, 566, 378 


\bibitem[\protect\citeauthoryear{{Lockman}}{{Lockman}}{1976}]{1976ApJ...209..4%
29L}
{Lockman} F.~J.,  1976, \apj, 209, 429

\bibitem[\protect\citeauthoryear{{Lockman}, {Blundell} \& {Goss}}{{Lockman}
  et~al.}{2007}]{2007MNRAS.381..881L}
{Lockman} F.~J.,  {Blundell} K.~M.,    {Goss} W.~M.,  2007, \mnras, 381, 881

\bibitem[Mezger 
\& Henderson(1967)]{1967ApJ...147..471M} Mezger, P.~G., \& Henderson, A.~P.\ 1967, \apj, 147, 471 

\bibitem[Miville-Desch{\^e}nes 
\& Lagache(2005)]{2005ApJS..157..302M} Miville-Desch{\^e}nes, M.-A., \& Lagache, G.\ 2005, \apjs, 157, 302

\bibitem[\protect\citeauthoryear{{Paladini}, {Burigana}, {Davies}, {Maino},
  {Bersanelli}, {Cappellini}, {Platania} \& {Smoot}}{{Paladini}
  et~al.}{2003}]{2003A&A...397..213P}
{Paladini} R.,  {Burigana} C.,  {Davies} R.~D.,  {Maino} D.,  {Bersanelli} M.,
  {Cappellini} B.,  {Platania} P.,    {Smoot} G.,  2003, \aap, 397, 213

\bibitem[\protect\citeauthoryear{{Paladini}, {Davies} \& {DeZotti}}{{Paladini}
  et~al.}{2004}]{2004MNRAS.347..237P}
{Paladini} R.,  {Davies} R.~D.,    {DeZotti} G.,  2004, \mnras, 347, 237

\bibitem[Putman et al.(2002)]{2002AJ....123..873P} Putman, M.~E., et al.\ 
2002, \aj, 123, 873 

\bibitem[\protect\citeauthoryear{{Reich}, {Fuerst}, {Haslam}, {Steffen} \&
  {Reif}}{{Reich} et~al.}{1984}]{1984A&AS...58..197R}
{Reich} W.,  {Fuerst} E.,  {Haslam} C.~G.~T.,  {Steffen} P.,    {Reif} K.,
  1984, \aaps, 58, 197

\bibitem[\protect\citeauthoryear{{Reynolds}}{{Reynolds}}{1985}]{1985ApJ...294.%
.256R}
{Reynolds} R.~J.,  1985, \apj, 294, 256

\bibitem[\protect\citeauthoryear{{Reynolds}}{{Reynolds}}{1991}]{1991ApJ...372L%
..17R}
{Reynolds} R.~J.,  1991, \apjl, 372, L17

\bibitem[\protect\citeauthoryear{{Rohlfs} \& {Wilson}}{{Rohlfs} \&
  {Wilson}}{2004}]{2004tra..book.....R}
{Rohlfs} K.,  {Wilson} T.~L.,  2004, {Tools of radio astronomy}

\bibitem[\protect\citeauthoryear{{Schlegel}, {Finkbeiner} \&
  {Davis}}{{Schlegel} et~al.}{1998}]{1998ApJ...500..525S}
{Schlegel} D.~J.,  {Finkbeiner} D.~P.,    {Davis} M.,  1998, \apj, 500, 525

\bibitem[\protect\citeauthoryear{{Shaver}, {McGee}, {Newton}, {Danks} \&
  {Pottasch}}{{Shaver} et~al.}{1983}]{1983MNRAS.204...53S}
{Shaver} P.~A.,  {McGee} R.~X.,  {Newton} L.~M.,  {Danks} A.~C.,    {Pottasch}
  S.~R.,  1983, \mnras, 204, 53

\bibitem[\protect\citeauthoryear{{Staveley-Smith}, {Wilson}, {Bird}, {Disney},
  {Ekers}, {Freeman}, {Haynes}, {Sinclair}, {Vaile}, {Webster} \&
  {Wright}}{{Staveley-Smith} et~al.}{1996}]{1996PASA...13..243S}
{Staveley-Smith} L.,  {Wilson} W.~E.,  {Bird} T.~S.,  {Disney} M.~J.,  {Ekers}
  R.~D.,  {Freeman} K.~C.,  {Haynes} R.~F.,  {Sinclair} M.~W.,  {Vaile} R.~A.,
  {Webster} R.~L.,    {Wright} A.~E.,  1996, Publications of the Astronomical
  Society of Australia, 13, 243

\bibitem[\protect\citeauthoryear{{Taylor} \& {Cordes}}{{Taylor} \&
  {Cordes}}{1993}]{1993ApJ...411..674T}
{Taylor} J.~H.,  {Cordes} J.~M.,  1993, \apj, 411, 674



\end{thebibliography}


\bsp 

\label{lastpage}

\end{document}